\documentclass[lettersize,journal]{IEEEtran}
\usepackage{amsmath,amsfonts}
\usepackage{algorithmic}
\usepackage{algorithm}
\usepackage{array}
\usepackage[caption=false,font=footnotesize,labelfont=rm,textfont=rm]{subfig}
\usepackage{textcomp}
\usepackage{placeins}
\usepackage{url}
\usepackage{verbatim}
\usepackage{graphicx}
\usepackage{cite}
\usepackage{amssymb}
\usepackage{setspace}
\usepackage{lineno} % 这个包就是latex自带的添加行号的包了
\usepackage{cases}
\usepackage{subfloat}
\usepackage{float}  %设置图片浮动位置的宏包
\usepackage{titlesec}
\usepackage{xcolor}
\usepackage[normalem]{ulem}

\usepackage{dblfloatfix}
\usepackage{needspace}

\begin{document} 
	
\bstctlcite{BSTcontrol}

\title{RIS-Assisted Multiuser NOMA Networks With Imperfect CSI Under Transceiver Hardware Impairments}

\author{Qian Zhang, Maoyuan Wang, Xuejun Cheng, Yufei Zhao, Guanghui Luo, Shiyao Guo

\thanks{Qian Zhang is with School of Computer and Communication Engineering, Northeastern University at Qinhuangdao, Qinhuangdao 066004, China (e-mail: zhangqian@neuq.edu.cn).}

\thanks{Maoyuan Wang, Xuejun Cheng, Shiyao Guo, and Qian Zhang are with School of Information Science and Engineering, Shandong University, Qingdao 266237, China (e-mail: maoyuanwang2024@mail.sdu.edu.cn, chengxuejun@mail.sdu.edu.cn, shiyaoguo@mail.sdu.edu.cn). }
	
%\thanks{The corresponding authors: Ju Liu; Hongji Xu. E-mail: \{juliu, hongjixu\}@sdu.edu.cn.}
\thanks{Yufei Zhao is with the School of Electrical and Electronics Engineering, Nanyang Technological University, Singapore 639798 (e-mail: yufei.zhao@ntu.edu.sg).}

\thanks{Guanghui Luo is with the National Mobile Communications Research Laboratory, Southeast University, Nanjing 210096, China (e-mail: guanghuiluo@seu.edu.cn). }
}
%This work was supported in part by the National Natural Science Foundation of China under Grant 62071275 and Grant 61901245; in part by the Key R\&D Plan of Shandong Province of China (Science and Technology Demonstration Project) under Grant 2021SFGC0701; in part by the Natural Science Foundation of Shandong Province of China under Grant ZR2020MF139.
%\markboth{SUBMITTED TO IEEE TRANSACTIONS ON VEHICULAR TECHNOLOGY, Jun. 2023.}%


\maketitle

\begin{abstract}
The incorporation of non-orthogonal multiple access (NOMA) and reconfigurable intelligent surface (RIS) provides a new solution to improve spectral efficiency. Unfortunately, the channel estimation error of RIS is noteworthy due to the absent active radio frequency link. Meanwhile, the transceivers at base stations (BS) and users also exist hardware impairments (HWI). The imperfect channel state information (CSI) and HWI will degrade the communication performance such as outage probability or data rate. To alleviate such performance deterioration, we propose a robust design scheme to minimize the BS transmit power when the target rate of users is satisfied in RIS-assisted single-beam and multi-beam NOMA networks with transceiver HWI and imperfect CSI. Further, the perfect successive interference cancellation (SIC) condition in single-beam and multi-beam NOMA networks is investigated, and the optimization problem is proposed based on perfect SIC conditions. To tackle the non-convex problem, we provide an effective algorithm combining Bernstein-type inequality and semi-definite relaxation. 
Numerical results demonstrate that the proposed scheme outperforms non-robust RIS-assisted NOMA scheme that ignores HWI and imperfect CSI.
Moreover, the multi-beam NOMA network obtains higher beam gain compared to the single-beam NOMA network, and thus there is a superior performance in multi-beam NOMA networks.

\end{abstract}


\vspace{-8pt}
\section{Introduction}

\IEEEPARstart{W}{ith} the widespread deployment of the fifth-generation (5G) wireless networks, academia and industry have gradually shifted their attention to the sixth-generation (6G) wireless networks \cite{8869705,9847080,8766143,9397776}. 6G cares for more emerging services and applications compared to 5G, which puts higher spectral efficiency (SE) requirements \cite{8766143,9397776,10772590}. Non-orthogonal multiple access (NOMA) can effectively enhance user fairness, improve SE, support massive connectivity, and reduce transmission latency, which is one of the most promising solutions in 6G networks \cite{9779790,9530717,11455905}. 
NOMA technology can effectively improve SE by exploiting the differences among user channels, but its performance is highly dependent on the propagation environment. In particular, blockage and outdoor-to-indoor (O2I) penetration loss pose significant coverage challenges for high-frequency communications \cite{10872772}. Reconfigurable intelligent surfaces (RISs) can reconfigure the propagation environment with low cost and low power consumption to improve the channel quality \cite{9779790,9530717,9765815, 1165892, 11139112,10446199}.
Therefore, it is natural to consider that the incorporation of RIS and NOMA can compensate for the weaknesses of NOMA to further improve SE and practicality. The combination of RIS and NOMA technology offers a promising multiple-access solution for future communications \cite{9779790,9530717}. In recent years, RIS-aided NOMA networks have emerged as a hot research topic, and relevant studies have emerged, such as \cite{9741332,9842326,10057422,9174801,wei2025mlnoma,zhang2024starris,ucargul2025risnoma}.

Unfortunately, due to the passive characteristic of RIS, the channel estimate of RIS is challenging \cite{9720945,ref16}. Thus far, installing active channel elements at the RIS is an effective method to obtain the CSI of each channel, but this approach inevitably increases the hardware cost, additional power consumption, and information exchange overhead \cite{ref18}. As a remedy, the beamforming design can be achieved to improve the communication performance when the CSI of the cascade channel between the base station (BS) and users is known. The cascade channel estimate is more tractable, but it may cause a larger channel estimation error (CEE). In fact, the perfect CSI is almost impossible because of the CEE, which inevitably renders the performance degradation of the studies based on the perfect CSI assumption. To alleviate such performance deterioration, it is necessary to investigate the robust transmission scheme in the RIS-assisted NOMA network with imperfect CSI. 
To the best of our knowledge, RIS-assisted NOMA networks under imperfect CSI have been less studied, and the most existing literature has merely studied RIS-assisted single-user or multi-user multiple-input single-output (MISO) networks under imperfect CSI \cite{ref18,ref19,ref21,ref22,zheng2025robust}. It is noteworthy that the difficulty of RIS channel estimation lies mainly in its absence of RF links. Then, the CSI of the direct channel between BS and users is a well-known problem and has many effective solutions given, but transceiver hardware impairments (HWI) that are introduced due to the presence of RF links are inevitable. Unlike additive white Gaussian noise (AWGN) at the receiver side, hardware impairments (HWI) generates distortion noise at the transmitter and receiver sides that is directly related to the transmit and receive signals \cite{ref23, 11455905}. Among the insightful observations, the network performance will degrade when the transceiver HWI in the RIS-assisted NOMA network is ignored. 
At present, several studies on RIS-assisted communication networks under transceiver hardware impairment have been carried out in the literature \cite{9842326,10057422,ref23,ref25,ref26,ref28}. These studies have in common either the study of single-user MISO scenarios \cite{ref23,ref25,ref26} or multi-user SISO scenarios \cite{ref28}, since the study under multi-user MISO systems with transceiver HWI is more complicated. In addition, there are fewer studies related to RIS-assisted NOMA networks in these studies, such as \cite{9842326,10057422,mead2025hwi,zhang2024starris,ucargul2025risnoma}.
The authors of \cite{9842326} analyzed the network performance of RIS-assisted NOMA systems under transceiver HWI.
The authors of \cite{10057422} investigated physical layer security (PLS) in two-user RIS-assisted NOMA networks under transceiver HWI through robust secure beamforming, while the authors of \cite{ref33} studied beamforming and jamming optimization for secure RIS-assisted NOMA networks. 
These studies provide useful insights into the decoding-rate relationship among different receivers, which motivates the subsequent SIC analysis.
In fact, in practical communications, there are almost always multiple users under multiple antenna BS. Therefore, it is very urgent to study the performance enhancement methods for RIS-assisted multi-user NOMA-MISO networks under transceiver HWI for future communications.

Moreover, robust beamforming under imperfect CSI has mainly been investigated through worst-case and probabilistic approaches. For bounded CSI errors, the S-procedure is commonly employed to transform worst-case QoS constraints into tractable linear matrix inequalities (LMIs), whereas statistical CSI errors are typically handled through outage-constrained formulations based on sphere bounding or Bernstein-type inequalities (BTIs) \cite{ref18,ref19,ref22,ref26}. Meanwhile, semi-definite relaxation (SDR), successive convex approximation (SCA), and alternating optimization (AO) have been widely applied to address rank-one constraints, non-convex terms, and the coupling between active and passive beamforming variables in RIS-assisted systems \cite{9741332,ref18,ref21}.
Various optimization methods have also been applied to RIS-assisted systems. For example, constrained stochastic successive convex approximation (CSSCA) has been employed for outage-constrained robust beamforming under imperfect CSI, while manifold optimization has been used to directly handle the unit-modulus constraints of RIS reflection coefficients \cite{zhao2021outage,elmossallamy2021ris}.
For the problem considered in this work, the statistical Gaussian CSI error model enables the BTI to transform the outage probability constraints into deterministic second-order cone (SOC) and linear matrix inequality (LMI) constraints, while the lifted beamforming variables lead to positive semidefinite matrices with rank-one constraints. Therefore, SDR and SCA are well suited to handling the rank-one constraints and the remaining non-convex terms, respectively. Meanwhile, AO is employed to decouple the active and passive beamforming variables for both the PCU and FCU scenarios.

%\blue{As mentioned above, both HWI and imperfect CSI are inevitable in practical communication systems. Nevertheless, to our knowledge, the joint robust transmission design for RIS-assisted multi-user NOMA networks simultaneously considering transceiver HWI and imperfect CSI remains insufficiently investigated.}
%\blue{Motivated by the above observations, we propose a robust transmission design scheme for RIS-assisted multi-user NOMA networks under imperfect CSI and transceiver HWI, aiming at minimizing the transmit power of the BS when the target rate of users is satisfied. }
%Also, it is known from past studies that NOMA is divided into two categories according to the number of beams: single-beam NOMA \cite{ref29} and multi-beam NOMA \cite{9741332}. Therefore, in this paper, we explore RIS-assisted single-beam and multi-beam NOMA networks and provide a detailed analysis of the two communication methods. The main contributions of our work are summarized as follows.

As mentioned above, both HWI and imperfect CSI are inevitable in practical communication systems. However, when they coexist in RIS-assisted multi-user NOMA systems, their impacts are not simply additive in terms of performance degradation. Specifically, transceiver HWI introduces distortion noise that is related to the transmit beamforming and received signal power, while imperfect CSI further introduces uncertainty into the corresponding decoding SINRs. Therefore, the SIC feasibility conditions, decoding order, and QoS constraints established under ideal hardware or perfect CSI cannot be directly applied to the considered system. Motivated by these observations, this work focuses on investigating the coupled effects of imperfect CSI and transceiver HWI on SIC and robust transmission, and establishes a corresponding transmit-power minimization framework for RIS-assisted multi-user NOMA networks.
In addition, NOMA can be categorized into single-beam NOMA \cite{ref29} and multi-beam NOMA \cite{9741332} according to the beamforming configuration. These two transmission modes exhibit different signal transmission and interference structures. In single-beam NOMA, all users share the same beamforming vector, whereas in multi-beam NOMA, the signals of different users are transmitted using their corresponding precoding vectors. Therefore, this work investigates both RIS-assisted single-beam and multi-beam NOMA networks, analyzes the SIC conditions for the two transmission modes, and compares their robust transmission performance under imperfect CSI and transceiver HWI. The main contributions of this work are summarized as follows.

\begin{itemize}
	
	%\item[$\bullet$] Both partial channel uncertainty (PCU) and full channel uncertainty (FCU) scenarios under the statistical CSI error model are considered to minimize the BS transmit power when the target rate of users is satisfied. Further, we give the outage probability constraint of the RIS-assisted multi-user NOMA network caused by imperfect CSI. Then, we derive a closed-form expression for the power of the distortion noise caused by the transceiver HWI in the multi-user network. As a result, combining HWI and imperfect CSI is explored to obtain more robust and superior performance in the RIS-assisted NOMA network.

	%\item[$\bullet$] The perfect successive interference cancellation (SIC) condition is explored based on {\it{Wyner}}'s theory of secure communication. As a result, we derive the equivalent-combined channel gain under single-cluster single-beam NOMA networks with imperfect CSI and transceiver HWI, and then the optimal decoding order for satisfying the perfect SIC in single-beam communication mode is obtained. In parallel, the available decoding method is proposed to guarantee successful SIC in the single-cluster multi-beam NOMA network.
	
	\item[$\bullet$]
	We establish a robust transmission model for RIS-assisted multi-user NOMA networks under transceiver HWI and statistical CSI uncertainty, considering both partial channel uncertainty (PCU) and full channel uncertainty (FCU) scenarios. We derive a closed-form expression for the distortion-noise power caused by transceiver HWI and incorporate it into the user SINRs, while the CSI uncertainty is characterized through outage-probability constraints. Based on this model, the joint effects of these practical impairments on the system performance can be further evaluated.
	
	\item[$\bullet$]
	We investigate the SIC conditions for single-beam and multi-beam NOMA under transceiver HWI and imperfect CSI. For single-beam NOMA, by jointly considering imperfect CSI and transceiver HWI, we derive the equivalent-combined channel gain and determine the user decoding order satisfying the SIC requirement. For multi-beam NOMA, since different precoding vectors make the decoding relationships among users more complicated, we further establish a corresponding decoding method to ensure successful SIC.

	\item[$\bullet$]
	An effective robust optimization framework is developed for solving the original non-convex problem. By exploiting the statistical Gaussian CSI error model, the outage probability constraints in both PCU and FCU scenarios are transformed into tractable deterministic constraints using the Bernstein-type inequality. Then, semi-definite relaxation (SDR) and successive convex approximation (SCA) are incorporated into an alternating optimization (AO) framework to handle the coupled variables and the associated non-convex constraints. In addition, a relaxation method together with an element-wise projection method is employed to obtain feasible solutions.
	
\end{itemize}

{\it{Notation:}} ${\cal{CN}}\left( {\boldsymbol{v}, \bold{V}}\right)$ is the circularly symmetric complex Gaussian (CSCG) distribution with mean vector $\boldsymbol{v}$ and covariance matrix $\bold{V}$. $\Vert\!\cdot\!\Vert$ and $\vert\!\cdot\!\vert$ denote the Euclidean norm of the vector and the absolute value of the scalar, respectively. $j$ stands for imaginary unit. $\mathbb{E}[\cdot]$ is the mean operator. $\mathbb{C}^{M \!\times\! N}$ represents the $M \!\times\! N$ complex matrix space, and $\bold{I}_N$ represents the $N \times N$ identity matrix. $\mathbb{H}^n$ indicates the space of $n\times n$ complex Hermitian matrices. $\bold{A} \!\succeq\! 0 $ denotes that matrix $\bold{A}$ is positive semi-definite. ${\rm{vec}}(\bold{A})$ stands for the column vectorization operation of the matrix $\bold{A}$. ${\rm{Tr}}(\bold{A})$, $\lambda(\bold{A})$ and $\lambda_{max}(\bold{A})$ represent the trace, eigenvalue of matrix $\bold{A}$, and maximum eigenvalue of matrix $\bold{A}$, respectively. $\bold{A}^H$ and $\widetilde{{\rm{diag}}}(\bold{A})$ represent the conjugate transposition of $\bold{A}$ and the diagonal matrix formed by its diagonal elements, respectively.

\section{System Model}
\subsection{RIS-assisted NOMA Network Architectures}

We study a downlink scenario in the RIS-assisted multi-user NOMA network under quasi-static flat-fading. As shown in Fig. 1, this downlink scenario consists of one BS with $M$ antennas, one RIS with $N$ reflecting elements, and $K$ single-antenna users. 
The direct BS-user links are obstructed and thus experience severe attenuation.
The RIS is deployed on a building close to users, which provides an unobstructed line-of-sight (LOS) path for users. The BS can send the superimposed signal to users through the direct or RIS reflection links. Meanwhile, the BS transmitting antenna and the user receiving antenna have some degree of impairment, thus generating distortion noise when sending or receiving signals. The NOMA user receives the BS signal and performs SIC to eliminate the high power interference from other users, ensuring the successful decoding of its signal.
\begin{figure*}[t]
	\centering
	\vspace{-0.5cm}
	\subfloat[single-cluster single-beam]{\includegraphics[width=2.8 in]{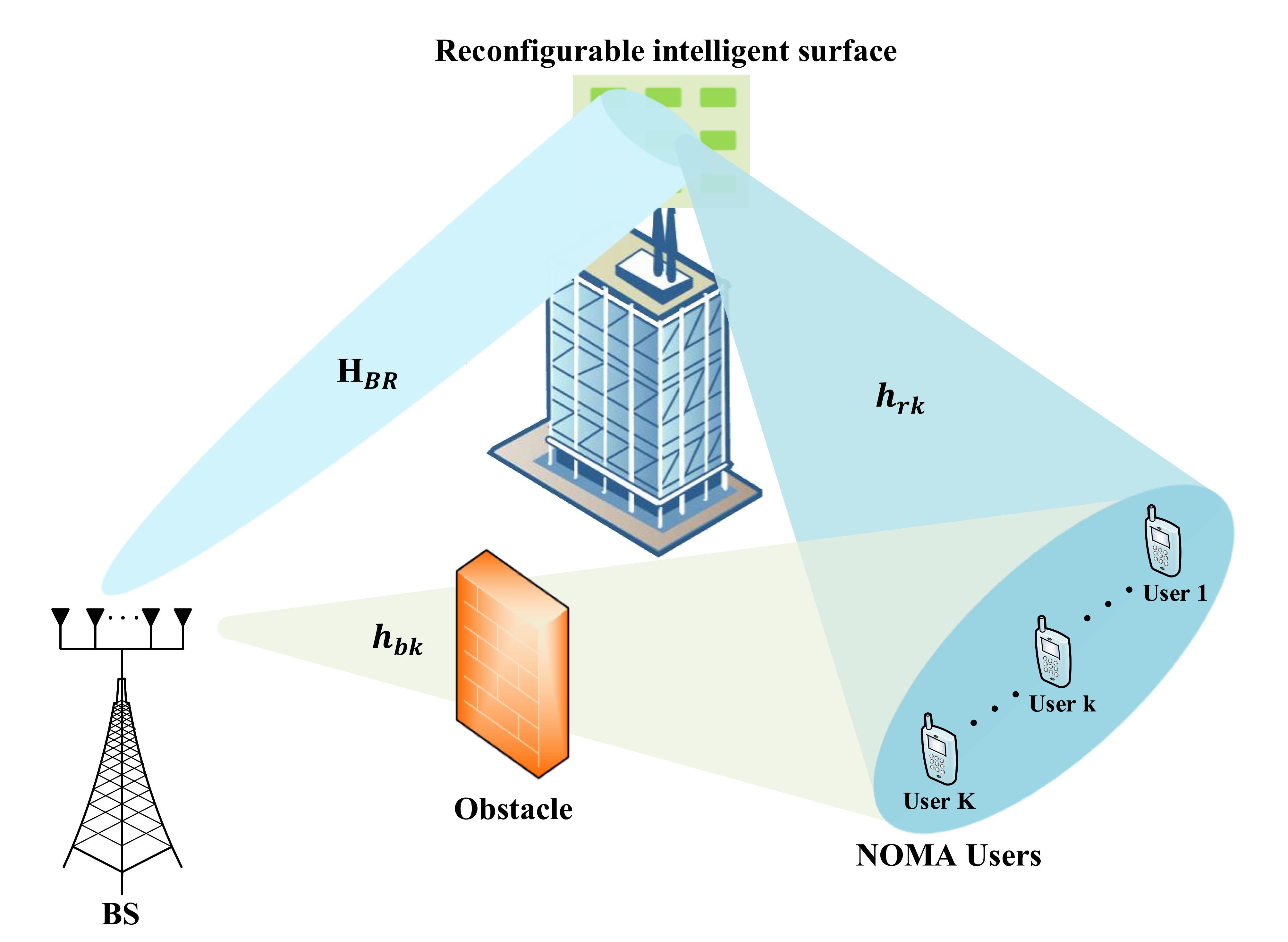}%
		\label{fig01_a}}
	\hfil
	\subfloat[single-cluster multi-beam]{\includegraphics[width=2.8 in]{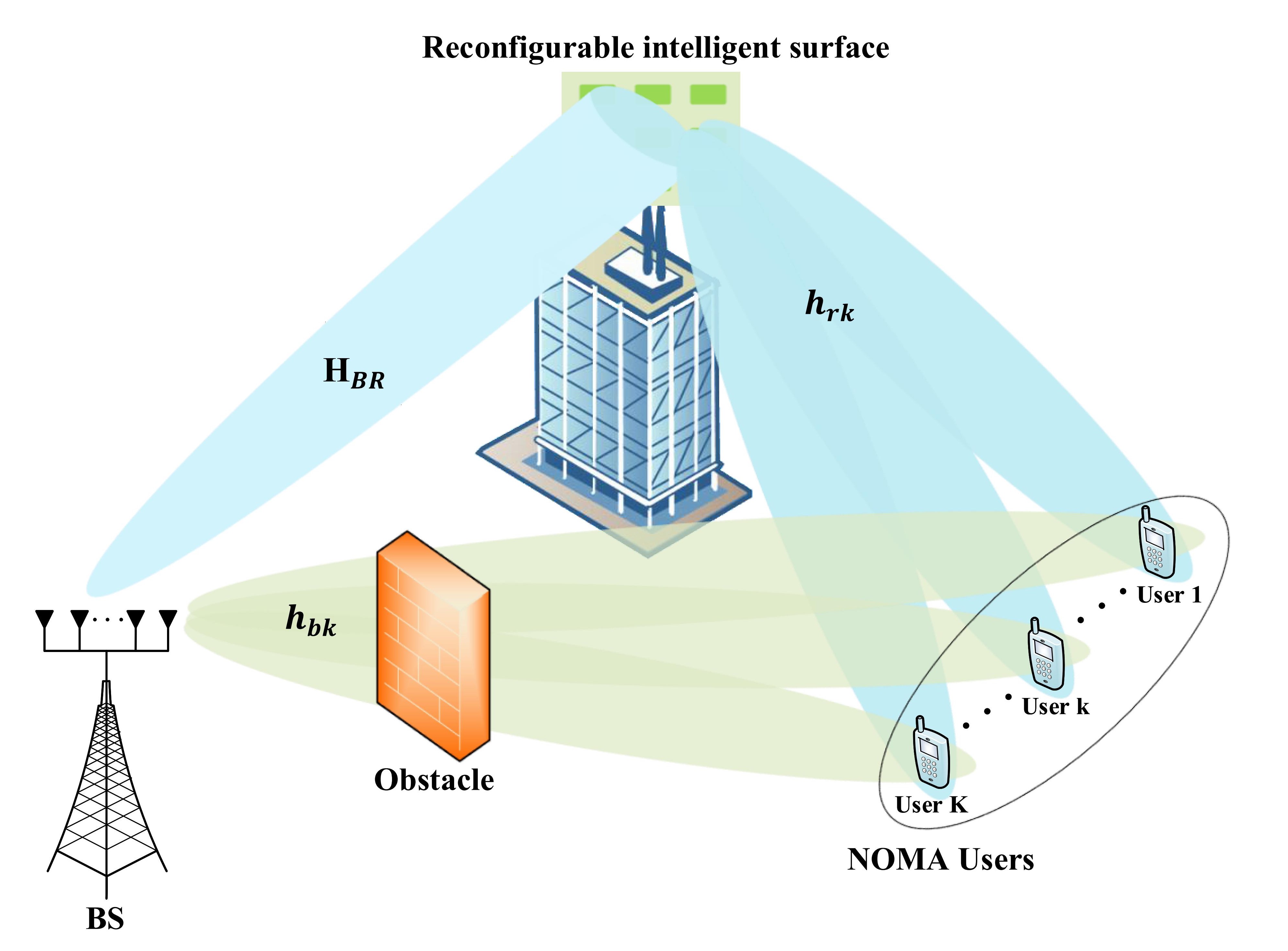}%
		\label{fig01_b}}
	\hfil
	\caption{The model of RIS-assisted multi-user NOMA network.}
	\label{fig01}
	\vspace{-10pt}
\end{figure*}

\vspace{-10pt}
\subsection{Channel Model and CSI Uncertainty}
\vspace{-3pt}

The wireless channels from the BS to the RIS, from the RIS to the $k$th user, and from the BS to the $k$th user are denoted by $\bold{H}_{BR}\in\mathbb{C}^{N \times M}$, $\boldsymbol{h}_{rk}\in\mathbb{C}^{N \times 1}$, and $\boldsymbol{h}_{k}\in\mathbb{C}^{M \times 1}$, respectively, where $\mathcal{K}=\{1,2,\dots,K\}$ denotes the user index set.
$\bold{H}_{BR}$ and $\boldsymbol{h}_{rk}$ are modeled as Rician channels \cite{9741332,9842326,10057422,9174801,ref21,ref25}.
\begin{equation*}
	\begin{split}
		\bold{H}_{BR} &= \sqrt{\beta_0 d_{BR}^{-\alpha_{BR}}} \left( {\sqrt{\frac{R_{BR}}{R_{BR}+1}} \bold{G}_L + \sqrt{\frac{1}{R_{BR}+1}} \bold{G}_S } \right), \\
		\boldsymbol{h}_{rk} &= \sqrt{\beta_0 d_{rk}^{-\alpha_{rk}}} \left( {\sqrt{\frac{R_{rk}}{R_{rk}+1}} \boldsymbol{g}_{Lk} + \sqrt{\frac{1}{R_{rk}+1}} \boldsymbol{g}_{Sk} } \right),
	\end{split}
\end{equation*}
where $k \in \mathcal{K}$ and $\beta_0$ represents the path loss at a one-meter reference distance. $d_{BR}$ and $d_{rk}$ denote the distance from the BS to the RIS, the RIS to the $k$th user, respectively. $R_{BR}>0$ and $R_{rk}>0$ are the fading factor of Rician distribution. $\alpha_{BR}$ and $\alpha_{rk}$ indicate the path-loss exponent. $\bold{G}_S \sim \mathcal{CN} \left( {\boldsymbol{0},\bold{I}} \right)$ and $\boldsymbol{g}_{Sk}\sim \mathcal{CN} \left( {\boldsymbol{0},\bold{I}} \right)$ express the channel scattering component. $\bold{G}_L \in \mathbb{C}^{N \times M}$ and $\boldsymbol{g}_{Lk} \in \mathbb{C}^{N \times 1}$ denote the deterministic LOS channel component.
Due to blockage, the BS-user channels $\boldsymbol h_k$ are modeled as NLOS Rayleigh fading channels \cite{ref39}.

When NOMA users decode the signal, the CSI of the corresponding channel needs to be obtained. However, the actual channel estimation will always have some degree of error, so we use a statistical CSI error model to quantify the CEE.
For a cascaded channel, satisfying $\{ \bold{H}_k = \overline{\bold{H}}_k + \Delta \bold{H}_k, \enspace {\rm{vec}} (\Delta \bold{H}_k) \sim {\cal{CN}} (\bold{0},\bold{\Gamma}_{H,k})\}$. For a direct channel, satisfying $\{\boldsymbol{h}_k = \overline{\boldsymbol{h}}_k + \Delta \boldsymbol{h}_k, \enspace \Delta \boldsymbol{h}_k \sim {\cal{CN}} (\bold{0},\bold{\Gamma}_{h,k})\}$. $\bold{\Gamma}_{H,k}$ and $\bold{\Gamma}_{h,k}$ are positive semi-definite error covariance matrices. For the convenience of the subsequent derivation, $\bold{\Gamma}_{H,k} = \phi_{H,k}^2 \bold{I}$ and $\bold{\Gamma}_{h,k} = \phi_{h,k}^2 \bold{I}$ are defined here \cite{ref18,ref21,ref22,ref26}. Therefore, the statistical CSI error can be rewritten as $\{{\rm{vec}} (\Delta \bold{H}_k) = \phi_{H,k} \bold{\Lambda}_{H,k},\enspace \bold{\Lambda}_{H,k} \sim \mathcal{CN} (\bold{0},\bold{I}), \enspace \forall k\in \mathcal{K}\}$ and $\{\Delta \boldsymbol{h}_k = \phi_{h,k} \bold{\Lambda}_{h,k},\enspace\bold{\Lambda}_{h,k} \sim \mathcal{CN} (\bold{0},\bold{I}), \enspace \forall k\in \mathcal{K}\}$.

We consider two uncertainty scenarios of channel \cite{ref22}, PCU and FCU. The PCU scenario includes inexact cascaded channel estimation and exact direct channel estimation, and the FCU scenario includes inaccurate cascaded channel estimation and direct channel estimation.
Based on these channel models, we next present the signal transmission models for the two NOMA communication modes.

\vspace{-8pt}
\subsection{Signal Transmission Model}
We assume that the BS transmits $K$ independent user-specific data symbols, denoted by $\boldsymbol{s}=[s_1,s_2,\ldots,s_K]^H$, where $s_k$ is the data symbol intended for the $k$th user. The data symbols obey the CSCG distribution and satisfy $\mathbb{E}[\boldsymbol{s}\boldsymbol{s}^H]=\bold{I}$.

\subsubsection{Single-Cluster Single-Beam Communication}
As shown in Fig.~1(a), the BS employs a common beamforming vector $\boldsymbol{w}_c$ for all NOMA users.
Accordingly, the BS transmits the superimposed signal as
\begin{equation}\label{x_1}
	{\mathop{\boldsymbol{x}}\nolimits}  = \sum_{k=1}^{K}\boldsymbol{w}_c \sqrt{\rho_k} s_k + \boldsymbol{\varrho}_t^c, 
\end{equation}
where $\rho_k$ denotes the power allocation factor of $k$th user such that $\sum_{k=1}^{K}\rho_k=1$, and $\boldsymbol{w}_c$ represents the precoding vector associated with a single-cluster. $\boldsymbol{\varrho}_t^c\sim{\cal{CN}}( 0,{\kappa_t}\widetilde{\textbf{{\rm{diag}}}} ( \boldsymbol{w}_c \boldsymbol{w}_c^H) )$ denotes transmit distortion noise caused by transmitter HWI, and $\kappa_t \in (0,1)$ is ratio between transmit distorted noise power and transmit signal power \cite{ref23,ref25,ref30,ref31}.
A common normalized impairment coefficient $\kappa_t$ is adopted to characterize the equivalent hardware quality of the BS RF chains, following the commonly adopted HWI model for multi-antenna transceivers \cite{ref26,6891254}.

\subsubsection{Single-Cluster Multi-Beam Communication}
As shown in Fig.~1(b), each BS beam characterized by the beamforming vector $\boldsymbol{w}_k$ carries the user-specific data symbol $s_k$. For each transmission mode, the RIS employs a common reflection matrix $\boldsymbol{\Theta}$ shared by all user data streams. Therefore, the BS transmits the superimposed signal as

\begin{equation}\label{x_2}
	{\mathop{\boldsymbol{x}}\nolimits}  = \sum_{k=1}^{K}\boldsymbol{w}_k s_k + \boldsymbol{\varrho}_t, 
\end{equation}
where ${{\mathop{\boldsymbol{w}}\nolimits} _k}\in\mathbb{C}^{M \times 1}$ is the precoding vector of the $k$th user. The transmit power at the BS is equal to $\sum_{k=1}^{K} \Vert \boldsymbol{w}_k \Vert^2$. 
$\boldsymbol{\varrho}_t\sim{\cal{CN}}( 0,{\kappa_t}\widetilde{\textbf{{\rm{diag}}}} ( \sum_{k=1}^{K}\boldsymbol{w}_k \boldsymbol{w}_k^H) )$ denotes transmit distortion noise caused by transmitter HWI.

For both communication modes, the squared norms of the corresponding beamforming vectors, i.e., $\boldsymbol{w}_c$ and $\{\boldsymbol{w}_k\}_{k=1}^{K}$, determine the transmit power, while their orientations determine the spatial beam directions. Therefore, optimizing these beamforming vectors also adjusts the power incident on the RIS through the BS-RIS link.

The signal received by the $k$th user from the BS is
\begin{equation}
	\label{y_k}
	\begin{split}
		y_k = \hat{y}_k + \varrho_k,\enspace \forall k\in \mathcal{K}, 
	\end{split}
\end{equation}
where $\hat{y}_k = (\boldsymbol{h}_{rk}^H \boldsymbol{\Theta} \bold{H}_{BR} + \boldsymbol{h}_{k}^H )\boldsymbol{x} + n_k = (\boldsymbol{v}^H \bold{H}_k + \boldsymbol{h}_{k}^H)\boldsymbol{x} + n_k$. $\bold{H}_k={\rm{diag}}(\boldsymbol{h}_{rk}^H) \bold{H}_{BR}$, $\mathcal{K}=\{1,2,\dots,K\}$. $\boldsymbol{\Theta}={\rm{diag}}(e^{j\theta_1},e^{j\theta_2},...,e^{j\theta_N})$, where $\theta_n \in [0,2\pi)$ is the reflecting signal phase variation through the $n$th RIS reflection element. $\boldsymbol{v}=\left[ {v_1,v_2,...v_N} \right]^H \in  \mathbb{C}^{N \times 1}$ represents the diagonal element of $\boldsymbol{\Theta}$. 
$n_k \sim \mathcal{CN}(0,\sigma_k^2)$ is AWGN.
$\varrho_k\sim {\cal{CN}}( {0,\kappa_r\mathbb{E}[|\hat{y}_k|^2]  } )$ denotes receive distortion noise caused by receiver HWI and $\kappa_r \in (0,1)$ stands for the ratio between received distorted noise and received undistorted noise \cite{ref23,ref25,ref30,ref31}.
The parameters $\kappa_t$ and $\kappa_r$ characterize the hardware impairment levels at the transmitter and receiver, respectively, and are closely related to the corresponding error vector magnitude (EVM) levels commonly used to evaluate practical transceiver quality. In general, an increase in $\kappa_t$ or $\kappa_r$ is accompanied by an increase in the corresponding EVM level, indicating more severe distortion introduced by the transceiver hardware \cite{6891254}.

\section{Problem Formulation}
In this section, we first characterize the SINR by incorporating the transceiver HWI, and then analyze the SIC conditions for the considered NOMA transmission modes. Based on these results, we formulate the robust transmit-power minimization problem under imperfect CSI.

\subsection{SINR Characterization}
In order to reveal the effect of HWI on the user-received signal-to-interference-plus-noise ratio (SINR), we derive a closed-form expression for the distortion noise power caused by the HWI to quantify it, as in \textbf{Proposition 1}.

\textbf{Proposition 1} \cite{10057422} \textbf{:} The power of distortion noise caused by receiver HWI in the multi-beam NOMA network is
\begin{equation}
	\label{E_yk_1}
	\begin{split}
		\mathbb{E}[\vert \hat{y}_k \vert^2] = (\boldsymbol{v}^H \bold{H}_k + \boldsymbol{h}_{k}^H) ( \sum_{k=1}^{K}\boldsymbol{w}_k \boldsymbol{w}_k^H + {\kappa_t}\widetilde{{\rm{diag}}} (\sum_{k=1}^{K}\boldsymbol{w}_k \boldsymbol{w}_k^H) ) \\ (\boldsymbol{h}_{k}+\bold{H}_k^H\boldsymbol{v}) + \sigma_k^2, \forall k \in \mathcal{K}.
	\end{split}
\end{equation}

Likewise, the power of distortion noise caused by receiver HWI in the single-beam NOMA network is 
\begin{equation}
	\label{E_yk_2}
	\begin{split}
		\mathbb{E}[\vert \hat{y}_k \vert^2] = (\boldsymbol{v}^H \bold{H}_k + \boldsymbol{h}_{k}^H) ( \boldsymbol{w}_c \boldsymbol{w}_c^H + {\kappa_t}\widetilde{{\rm{diag}}} (\boldsymbol{w}_c \boldsymbol{w}_c^H) ) \\ (\boldsymbol{h}_{k}+\bold{H}_k^H\boldsymbol{v}) + \sigma_k^2, \forall k \in \mathcal{K}.
	\end{split}
\end{equation}

{\it{Proof:}} See Appendix A.	\hfill$\blacksquare$

These distortion-noise expressions are incorporated into the SINR characterization for the two NOMA transmission modes.

\subsubsection{Single-Cluster Single-Beam Communication}
Based on the single-beam signal model in \eqref{x_1}, the corresponding decoding SINRs are given by
\begin{equation}
	\label{SINR_1}
	\begin{split}
		\gamma_k^k &= \frac{\vert \widetilde{\bold{H}}_k\boldsymbol{w}_c  \vert^2 \rho_k}{ \widetilde{\bold{H}}_k  (\sum_{i=k+1}^{K} \rho_i\boldsymbol{w}_c \boldsymbol{w}_c^H  + \bold{\Psi}_c) \widetilde{\bold{H}}_k^H + \widetilde{\sigma}_k^2 }, \\
		\gamma_l^k &= \frac{\vert \widetilde{\bold{H}}_l \boldsymbol{w}_c  \vert^2 \rho_k}{\widetilde{\bold{H}}_l (\sum_{i=k+1}^{K} \rho_i\boldsymbol{w}_c \boldsymbol{w}_c^H + \bold{\Psi}_c)\widetilde{\bold{H}}_l^H + \widetilde{\sigma}_l^2 }, \\
		\gamma_K^K &= \frac{\vert \widetilde{\bold{H}}_K \boldsymbol{w}_c  \vert^2 \rho_K}{ \widetilde{\bold{H}}_K \bold{\Psi}_c\widetilde{\bold{H}}_K^H + \widetilde{\sigma}_K^2 }, \enspace \forall k \in \mathcal{S},\forall l \in \mathcal{T},
	\end{split}
\end{equation}
where $\widetilde{\bold{H}}_k = \boldsymbol{v}^H \bold{H}_k + \boldsymbol{h}_{k}^H$, $\widetilde{\sigma}_k^2 = (1+\kappa_r)\sigma_k^2$, $\mathcal{S}=\{1,2, ... ,K-1\}$, $\mathcal{T}=\{k,k+1, ... ,K\}$, and $\bold{\Psi}_c = \kappa_r \boldsymbol{w}_c \boldsymbol{w}_c^H + {(1+\kappa_r)\kappa_t}\widetilde{{\rm{diag}}} (\boldsymbol{w}_c \boldsymbol{w}_c^H)$.

\subsubsection{Single-Cluster Multi-Beam Communication}
Based on the multi-beam signal model in \eqref{x_2}, the SINR for the $k$th user to decode its own signal is given by
\begin{equation}
	\label{SINR_2}
	\begin{split}
		\gamma_k^k = \frac{\vert \widetilde{\bold{H}}_k \boldsymbol{w}_k  \vert^2}{\widetilde{\bold{H}}_k (\sum_{i=k+1}^{K}\boldsymbol{w}_i \boldsymbol{w}_i^H + \bold{\Psi})\widetilde{\bold{H}}_k^H + \widetilde{\sigma}_k^2 }, \enspace \forall k \in \mathcal{S},
	\end{split}
\end{equation}
where 
$$
\bold{\Psi} = \kappa_r\sum_{k=1}^{K}\boldsymbol{w}_k \boldsymbol{w}_k^H + {(1+\kappa_r)\kappa_t}\widetilde{{\rm{diag}}} (\sum_{k=1}^{K}\boldsymbol{w}_k \boldsymbol{w}_k^H).
$$

Further, the corresponding SINR at the $l$th user for decoding the $k$th user's signal is given by
\begin{equation}
	\label{SINR_3}
	\begin{split}
		\gamma_l^k = \frac{\vert \widetilde{\bold{H}}_l \boldsymbol{w}_k  \vert^2}{\widetilde{\bold{H}}_l (\sum_{i=k+1}^{K}\boldsymbol{w}_i \boldsymbol{w}_i^H \!+\! \bold{\Psi})\widetilde{\bold{H}}_l^H \!+\! \widetilde{\sigma}_l^2 }, \forall k \!\in\! \mathcal{S},\forall l \!\in\! \mathcal{T}.
	\end{split}
\end{equation}

For the $K$th user, the corresponding decoding SINR is
\begin{equation}
	\label{SINR_4}
	\begin{split}
		\gamma_K^K = \frac{\vert \widetilde{\bold{H}}_K \boldsymbol{w}_K  \vert^2}{ \widetilde{\bold{H}}_K \bold{\Psi}\widetilde{\bold{H}}_K^H + \widetilde{\sigma}_K^2 }.
	\end{split}
\end{equation}

\subsection{Perfect SIC Condition}
It is essential to guarantee perfect SIC, which is fundamental to NOMA networks \cite{ref29}.
Inspired by {\it{Wyner}}'s secure communication theory \cite{ref32}, particularly the comparison of decoding rates among different receivers, we characterize the SIC condition based on the decoding-rate relationship among NOMA users.
Specifically, the eavesdropper will not be able to decode the legitimate user signal if BS transmits legitimate signal at the secrecy rate $C_e = [\mathcal{R} - \mathcal{R}_e]^+$, where $\mathcal{R}$ denotes the legitimate user rate, $\mathcal{R}_e$ denotes the eavesdropping rate, and $[x]^+=max(0,x)$. Therefore, to ensure successful SIC in the NOMA network, the condition $\gamma_l^k\geq \gamma_k^k, \enspace \forall k \in \mathcal{S},\forall l \in \mathcal{T}$ should be satisfied. Accordingly, we define an equivalent-combined channel gain to determine the optimal decoding order and guarantee successful SIC.

\textbf{Theorem 1:} In the single-beam communication scenario, defining $G_k = \frac{\vert (\boldsymbol{v}^H \bold{H}_k + \boldsymbol{h}_{k}^H)\boldsymbol{w}_c  \vert^2 \rho_k}{ \bold{\Psi}_c^e }, \enspace \forall k \in \mathcal{K}$ as the equivalent-combined channel gain for the $k$th user, where $\bold{\Psi}_c^e = (1+\kappa_r)\kappa_t(\boldsymbol{v}^H \bold{H}_k + \boldsymbol{h}_{k}^H)  \widetilde{{\rm{diag}}} (\boldsymbol{w}_c \boldsymbol{w}_c^H) (\boldsymbol{h}_{k}+\bold{H}_k^H\boldsymbol{v}) + (1+\kappa_r)\sigma_k^2$. Then, the channel gain order satisfying $G_1 \leq G_2 \leq \dots \leq G_K$ is the optimal decoding order. 

{\it{Proof:}} See Appendix B.	\hfill$\blacksquare$ 

In the single-beam communication scenario, the perfect SIC condition can be readily satisfied using SDR and SCA, and the optimal $\rho_k^{opt}$ has been given in \cite{ref29}. However, due to the area broadcast characteristics of single-beam communications, the energy efficiency and security performance of single-beam communications are lower than those of multi-beam communications. 

Following \textbf{Theorem 1}, the perfect SIC condition for the multi-beam communication scenario can be derived as follows.
\begin{equation} 
	\label{H_k} 
	\begin{split} 
		\frac{\vert \widetilde{\bold{H}}_k \boldsymbol{w}_k  \vert^2}{\vert \widetilde{\bold{H}}_l \boldsymbol{w}_k  \vert^2} \overset{(a)}{\leq} \frac{\widetilde{\bold{H}}_k(\sum_{i=k+1}^{K}\boldsymbol{w}_i \boldsymbol{w}_i^H + \bold{\Psi})\widetilde{\bold{H}}_k^H + \vartheta}{\widetilde{\bold{H}}_l (\sum_{i=k+1}^{K}\boldsymbol{w}_i \boldsymbol{w}_i^H + \bold{\Psi})\widetilde{\bold{H}}_l^H + \vartheta}  \overset{(b)}{\leq} 1, 
	\end{split}  
\end{equation} 
where $(b)$ holds due to the stronger channel gain of the $l$th user, and $\vartheta = (1+\kappa_r)\sigma^2$. 

The condition $\gamma_l^k\geq \gamma_k^k$ is difficult to satisfy because of inequality $(b)$. To address this issue, we can set $\xi \leq {\rm{min}}\{  \frac{\widetilde{\bold{H}}_k (\sum_{i=k+1}^{K}\boldsymbol{w}_i \boldsymbol{w}_i^H + \bold{\Psi})\widetilde{\bold{H}}_k^H + \widetilde{\sigma}_k^2}{\widetilde{\bold{H}}_l (\sum_{i=k+1}^{K}\boldsymbol{w}_i \boldsymbol{w}_i^H + \bold{\Psi})\widetilde{\bold{H}}_l^H + \widetilde{\sigma}_l^2} \}$ and $ \vert \widetilde{\bold{H}}_k \boldsymbol{w}_k  \vert^2 \leq \xi \vert \widetilde{\bold{H}}_j \boldsymbol{w}_k  \vert^2 $ to guarantee the perfect SIC. However, the two inequalities are strongly coupled and difficult to optimize, so this method may not always match the appropriate beamforming. A simple method satisfying the perfect SIC is to set $\gamma_l^k$ greater than the threshold $\gamma_{th}$, which requires a reasonable setting of the threshold $\gamma_{th}$.

In addition, condition $\gamma_l^k\geq \gamma_k^k$ may lead to a situation where the $k$th user cannot decode the $k$th user signal, especially in secure communication scenarios. As a result, the revised condition $\gamma_l^k = \gamma_k^k$ must be satisfied for successful communication in NOMA networks. Nevertheless, this revised condition is overly stringent and difficult to satisfy in all cases. In general, we consider that the SIC can be successfully executed and communicated when $\gamma_l^k$ and $\gamma_k^k$ are sufficiently close, which can satisfy the communication needs in most scenarios. For the considered RIS-assisted single-cluster multi-beam NOMA networks under HWI and imperfect CSI, successful SIC can be ensured by requiring $\gamma_k^{tar}$ to be greater than the SIC decoding threshold. Specifically, each $\gamma_l^k,\forall k\in \mathcal{S}, \forall l \in \mathcal{T}$ should be no smaller than the target SINR $\gamma_k^{tar},\forall k\in \mathcal{S}$ of the $k$th user to ensure performing the SIC successfully. Therefore, condition $\gamma_l^k \geq \gamma_k^{tar}, \forall k\in \mathcal{S}, \forall l \in \mathcal{T}$ must be satisfied \cite{ref33,ref34,ref35}. $\gamma_k^{tar}$ can be expressed as $\gamma_k^{tar} = {\rm{min}}\{ \gamma_k^k,\gamma_{k+1}^k,\dots,\gamma_K^k \}$. In addition, at the optimum, both $\gamma_l^k$ and $\gamma_k^{tar}$ are typically close to or even equal to the threshold, thereby guaranteeing successful SIC and communication.
It should be noted that when the decoding SINR approaches the target threshold, the distortion noise caused by HWI and the CSI uncertainty affect the SIC decoding performance. In this work, these effects are incorporated into the robust design framework, where the HWI-induced residual interference is included in the equivalent interference term $\Psi$ of the SINR expression, and the CSI uncertainty is characterized by the statistical channel error model and handled through the outage probability constraints. Therefore, considering HWI and CSI uncertainty, the robust framework can guarantee that the SIC decoding SINR satisfies the predefined target threshold from a statistical perspective

Accordingly, the decoding rate of the $k$th user signal is
\begin{equation} 
	\label{R_k} 
	\begin{split} 
		\mathcal{R}_k &= {\rm{log_2}}\,(1+\gamma_k^{tar}) = {\rm{log_2}}\,(1+ \mathop{{\rm{min}}}\limits_{l\in \mathcal{T}} \gamma_l^k), \enspace \forall k\in \mathcal{S}. 
	\end{split} 
\end{equation}

\vspace{-22pt}
\subsection{Power Minimization}
\vspace{-6pt}
Based on the above SINR characterization and SIC conditions, we aim to minimize the BS transmit power while satisfying the target-rate requirements of users under multi-beam communication. In addition, the algorithm can directly solve the transmit power minimization problem in the single-cluster single-beam communication scenario under optimal power allocation factor $\rho_k^{opt}$. Specifically, due to imperfect CSI, the outage probability constraint is introduced into the resulting optimization problem as follows:
\begin{subequations} 
	\label{P1} 
	\begin{align} 
		\bold{P1}: 
		\mathop{{\rm min}}\limits_{\{\boldsymbol{w}_k\}_{k=1}^K,\boldsymbol{v}} \  
		&\sum_{k=1}^{K}\boldsymbol{w}_k^H\boldsymbol{w}_k 
		\label{P1a}\\ 
		{\rm s.t.}\quad 
		&{\rm Pr}\left\{\mathcal{R}_k\geq R_{th}^k\right\} 
		\geq 1-P_{out}^k, 
		\forall k\in\mathcal{K} 
		\label{P1b}\\ 
		&\left\Vert\boldsymbol{w}_{j+1}\right\Vert^2 
		\leq 
		\left\Vert\boldsymbol{w}_{j}\right\Vert^2, 
		\forall j\in\mathcal{S} 
		\label{P1c}\\ 
		&|v_n|=1,\quad n=1,2,\ldots,N 
		\label{P1d} 
	\end{align} 
\end{subequations} 

\noindent where $R_{th}^k$ denotes the rate threshold for decoding the $k$th user signal, and $P_{out}^k$ expresses the outage probability when decoding the $k$th user signal. 
As defined in (11), $R_k$ is determined by the minimum decoding SINR among the users required to decode the $k$th user signal. Therefore, constraint~\eqref{P1b} guarantees that the required decoding rate is satisfied with a probability of at least $1-P_{\mathrm{out}}^k$ under imperfect CSI, providing the corresponding SIC decoding reliability. Constraint~\eqref{P1c} imposes the transmit-power ordering $\|\mathbf{w}_{j+1}\|^2 \leq \|\mathbf{w}_{j}\|^2$, which is consistent with the adopted SIC decoding order and facilitates successive interference cancellation. Constraint \eqref{P1d} represents the unit-modulus constraint of the RIS reflection coefficients. 

According to \cite{ref36}, $\bold{P1}$ is a challenging non-convex problem and is difficult to solve directly. First, the outage-probability constraint~\eqref{P1b} involves random CSI errors and does not admit a directly tractable deterministic form. Moreover, the precoding	vector $\{\boldsymbol{w}_k\}_{k=1}^{K}$ and the RIS reflection vector $\boldsymbol{v}$ are deeply coupled. Therefore, we adopt an alternating optimization framework and decompose $\bold{P1}$ into two sub-problems: the active beamforming problem $\bold{P2}$ with fixed $\boldsymbol{v}$ and the passive beamforming problem $\bold{P3}$ with fixed $\{\boldsymbol{w}_k\}_{k=1}^{K}$.
\begin{subequations} 
	\label{P2} 
	\begin{align} 
		\bold{P2}:	\mathop{{\rm{min}}}\limits_{\{\boldsymbol{w}_k\}_{k=1}^K} \quad & 
		\sum_{k=1}^{K}\boldsymbol{w}_k^H \boldsymbol{w}_k \label{P2a} \\ 
		s.t. \quad& 
		{\rm{Pr}} \{ \mathcal{R}_k \!\geq\! R_{th}^k \} \!\geq\! 1\!-\!P_{out}^k, \forall k \!\in\! \mathcal{K}, \label{P2b} \\ 
		&\Vert \boldsymbol{w}_{j+1} \Vert^2 \leq \Vert \boldsymbol{w}_{j} \Vert^2, \enspace \forall j\in \mathcal{S}. \label{P2c} 
	\end{align} 
\end{subequations} 
\vspace{-18pt}
\begin{subequations} 
	\label{P3} 
	\begin{align} 
		\bold{P3}:	\mathop{{\rm{Find}}} \quad & 
		\boldsymbol{v} \label{P3a} \\ 
		s.t. \quad& 
		{\rm{Pr}} \{ \mathcal{R}_k \geq R_{th}^k \} \geq 1-P_{out}^k, \enspace \forall k\in \mathcal{K}, \label{P3b} \\ 
		& \vert v_n \vert=1, \enspace n=1,2,...N.  \label{P3c} 
	\end{align} 
\end{subequations}

\vspace{-18pt}
\section{Robust Beamforming Optimization Solution}
\vspace{-8pt}

In this section, we provide an efficient solution for $\bold{P2}$ and $\bold{P3}$. Specifically, the {\it{Bernstein-type Inequality}} is employed to solve constraints~\eqref{P2b} and~\eqref{P3b}. Meanwhile, the SDR technique and the SCA algorithm are used to convert the non-convex problem to a convex problem. In the end, the optimal solution is obtained using the alternate optimization (AO) algorithm. Then, the {\it{Bernstein-type Inequality}} is given as follows.

\textbf{Lemma 1:} ({\it{Bernstein-type Inequality}} \cite{ref19}) Assume $\bold{Q}_k\in \mathbb{H}^n$, $\boldsymbol{q}_k \in \mathbb{C}^{n \times 1}$, $c_k\in \mathbb{R}$, $\boldsymbol{z}_k\in \mathbb{C}^{n \times 1}$, $\boldsymbol{z}_k \sim \mathcal{CN}(\bold{0},\bold{I})$, slack variables $\boldsymbol{x} = [x_1,x_2,...,x_k,...,x_K]^T$ and $\boldsymbol{y} = [y_1,y_2,...,y_k,...,y_K]^T$, for $\forall p_k\in [0,1]$, we can obtain
\begin{subequations} 
	\label{Lemma_1} 
	\begin{align} 
		&{\rm Pr}\!\left\{
		\boldsymbol{z}_k^H \bold{Q}_k \boldsymbol{z}_k
		\!+\!2{\rm Re}\!\left\{\boldsymbol{z}_k^H\boldsymbol{q}_k\right\}
		\!+\!c_k \!\geq\!0
		\right\}
		\geq1-p_k,
		\label{Lemma_1a}
		\\
		\Rightarrow \enspace &
		{\rm Tr}\{\bold{Q}_k\}
		-\sqrt{2{\rm log}(1/p_k)}
		\sqrt{\Vert\bold{Q}_k\Vert_F^2
			+2\Vert\boldsymbol{q}_k\Vert_2^2}
		\nonumber
		\\
		&
		+{\rm log}(p_k)\lambda^+(\bold{Q}_k)
		+c_k\geq0,\enspace
		\forall k\in\mathcal{K},
		\label{Lemma_1b}
		\\
		\Rightarrow \enspace &
		\left\{
		\begin{aligned}
			&{\rm Tr}\{\bold{Q}_k\}
			-\sqrt{2{\rm log}(1/p_k)}x_k
			+{\rm log}(p_k)y_k+c_k\geq0,
			\\
			&\left\|
			[{\rm vec}(\bold{Q}_k),
			\sqrt{2}\boldsymbol{q}_k]^T
			\right\|
			\leq x_k,
			\\
			&y_k\bold{I}+\bold{Q}_k\succeq0,\enspace
			y_k\geq0,\enspace
			\forall k\in\mathcal{K}.
		\end{aligned}
		\right.
		\label{Lemma_1c}
		\raisetag{0.8\baselineskip}
	\end{align} 
\end{subequations}

where $\lambda^+(\bold{Q}_k)
=\max\{\lambda_{\max}(-\bold{Q}_k),0\}$.

Since $\mathop{{\rm{min}}}\limits_{l\in \mathcal{T}} \gamma_l^k$ cannot be determined, the following inequality is captured to ensure~\eqref{P1b}.
\begin{equation}
	\label{Pr_1}
	\begin{split}
		& {\rm{Pr}} \{ {\rm{log_2}}\,(1\!+\! \gamma_K^K) \geq R_{th}^K \} \!\geq\! 1\!-\! P_{out}^K,\\
		&{\rm{Pr}} \{ {\rm{log_2}}\,(1\!+\!\gamma_l^k) \geq R_{th}^k \} \!\geq\! 1\!-\! P_{out}^k, \enspace \forall k\in \mathcal{S}, \forall l \in \mathcal{T}.
	\end{split}
\end{equation}

When \eqref{P1b} is satisfied, \eqref{Pr_1} must hold, and conversely, when \eqref{Pr_1} is satisfied, \eqref{P1b} must also hold. Therefore, \eqref{Pr_1} is a sufficient condition for \eqref{P1b}, and the two are equivalent.

To solve~\eqref{Pr_1}, we perform the following transformation.
\begin{equation} 
	\label{Pr_2} 
	\eqref{Pr_1}\Rightarrow 
	\left\{ 
	\begin{aligned} 
		& \Pr\left\{\widetilde{\mathbf H}_{l}\boldsymbol{\Phi}_{k}\widetilde{\mathbf H}_{l}^{H} 
		-\widetilde{\sigma}_{l}^{2}\geq 0\right\}\geq 1-P_{\mathrm{out}}^{k},\\ 
		& \Pr\left\{\widetilde{\mathbf H}_{K}\boldsymbol{\Phi}_{K}\widetilde{\mathbf H}_{K}^{H} 
		-\widetilde{\sigma}_{K}^{2}\geq 0\right\}\geq 1-P_{\mathrm{out}}^{K}, 
	\end{aligned} 
	\right. 
\end{equation} 
where $\forall k \in \mathcal{S}$, $\forall l \in \mathcal{T}$, 
$\boldsymbol{\Phi}_k=
\frac{\boldsymbol{w}_k \boldsymbol{w}_k^H}{2^{R_{\mathrm{th}}^k}-1}
-\sum_{i=k+1}^{K}\boldsymbol{w}_i\boldsymbol{w}_i^H
-\boldsymbol{\Psi}$, 
and 
$\boldsymbol{\Phi}_K=
\frac{\boldsymbol{w}_K \boldsymbol{w}_K^H}{2^{R_{\mathrm{th}}^K}-1}
-\boldsymbol{\Psi}$.

Then, the {\it{Kronecker Products}} and {\it{vectorization}} of the matrix are introduced to convert~\eqref{Pr_2} into the form of~\eqref{Lemma_1a}, and the outage probability constraint can be solved using {\it{Bernstein-type Inequality}}. 

\subsection{PCU Scenario}
\textbf{Proposition 2:} Based on the PCU channel model in Section II-B and the matrix properties in \cite{ref38,ref39}, by defining $\boldsymbol{V}=\boldsymbol{v}\boldsymbol{v}^{H}$, inequality~\eqref{Pr_2} can be reformulated as
\begin{equation}
	\label{eq:pcu-chance}
	\begin{split}
		{\rm{Pr}} \{  \bold{\Lambda}_{H,l}^H \bold{Q}_{l,k} \bold{\Lambda}_{H,l} + 2Re\{ \bold{\Lambda}_{H,l}^H \boldsymbol{r}_{l,k} \} + c_{l,k}  \geq 0 \} \geq 1-P_{out}^k,
	\end{split}
\end{equation}
where  $\forall k\in \mathcal{K}, \forall l\in \mathcal{T},$
\begin{equation}
	\label{eq:pcu-definitions}
	\begin{split}
		\bold{Q}_{l,k} &\!=\! \phi_{H,l}^2 (\bold{\Phi}_k^H \!\!\otimes\!\! \bold{V}), 
		\boldsymbol{r}_{l,k} \!=\! \phi_{H,l} {\rm{vec}}(\!(\bold{V}\overline{\bold{H}}_l \!+\!\boldsymbol{v}\boldsymbol{h}_l^H\!)\bold{\Phi}_k\!),  \\
		c_{l,k} &\!=\! (\boldsymbol{v}^H \overline{\bold{H}}_l \!+\! \boldsymbol{h}_l^H) \bold{\Phi}_k (\boldsymbol{h}_l \!+\! \overline{\bold{H}}_l^H \boldsymbol{v} ) \!-\! (1 \!+\!\kappa_r)\sigma_k^2.
	\end{split}
\end{equation}

{\it{Proof:}} See Appendix C.	\hfill$\blacksquare$

Further, to solve~\eqref{eq:pcu-chance}, the auxiliary variables $\boldsymbol{x} = [x_{1,1},x_{2,1},...,x_{l,k},...,x_{K,K}]^H, \enspace \forall k\in \mathcal{K}, \forall l\in \mathcal{T}$ and $\boldsymbol{y} = [y_{1,1},y_{2,1},...,y_{l,k},...,y_{K,K}]^H, \enspace \forall k\in \mathcal{K}, \forall l\in \mathcal{T}$ are introduced. Then, applying \textbf{Lemma 1}, the constraint~\eqref{eq:pcu-chance} can be converted into the deterministic form as 
\begin{equation}
	\label{eq:pcu-deterministic}
	\left\{
	\begin{aligned}
		& {\rm{Tr}}\{\bold{Q}_{l,k}\}-\sqrt{2{\rm{log}}\,(1/P_{out}^k)}x_{l,k}+{\rm{log}}\,(P_{out}^k)y_{l,k}+c_{l,k}\geq 0, \\
		& \sqrt{ \Vert \bold{Q}_{l,k} \Vert_F^2 +2\Vert \boldsymbol{r}_{l,k} \Vert_2^2 } \leq x_{l,k}, \enspace
		y_{l,k} \bold{I} + \bold{Q}_{l,k} \succeq 0, \enspace y_{l,k}\geq 0.
	\end{aligned}
	\right.
\end{equation}

Based on matrix properties \cite{ref37,ref38}, the components in~\eqref{eq:pcu-deterministic} can be simplified as follows
\begin{equation}
	\label{eq:pcu-components}
	\begin{split}
		{\rm{Tr}}\{\bold{Q}_{l,k}\} v&= \phi_{H,l}^2 {\rm{Tr}}\{\bold{\Phi}_k \otimes \bold{V}\} =\phi_{H,l}^2N{\rm{Tr}}\{\bold{\Phi}_k\}, \\
		\Vert \bold{Q}_{l,k} \Vert_F^2 &= \phi_{H,l}^4 \Vert \bold{\Phi}_k \otimes \bold{V} \Vert_F^2  = \phi_{H,l}^4 N^2 \Vert \bold{\Phi}_k \Vert_F^2, \\
		 \Vert \boldsymbol{r}_{l,k} \Vert_F^2& = \phi_{H,l}^2 N \Vert ((\boldsymbol{v}^H\overline{\bold{H}}_l+\boldsymbol{h}_l^H)\bold{\Phi}_k) \Vert_2^2.
	\end{split}
\end{equation}

\subsubsection{Active Beamforming Optimization}
Problem $\bold{P2}$ is to optimize $\{\boldsymbol{w}_k\}_{k=1}^K$ at a fixed  $\boldsymbol{v}$ and is solved by the SDR technique. Defining $\bold{W}_k = \boldsymbol{w}_k \boldsymbol{w}_k^H$, based on the above derivation, problem $\bold{P2}$ is approximated as
\begin{subequations}
	\label{P4}
	\begin{align}
		\bold{P4}:	&\mathop{{\rm{min}}}\limits_{\{\bold{W}_k\}_{k=1}^K,\boldsymbol{x},\boldsymbol{y}} \quad 
		\sum_{k=1}^{K}{\rm{Tr}}\{\bold{W}_k\} \label{P4a} \\
		&\enspace \enspace s.t. \quad\enspace
		\phi_{H,l}^2 N {\rm{Tr}}\{\bold{\Phi}_k\} - \sqrt{2 {\rm{log}}\,(1/P_{out}^k)}x_{l,k} \nonumber \\  &\qquad\qquad\qquad\qquad  +{\rm{log}}\,(P_{out}^k)y_{l,k} +c_{l,k}\geq 0, \label{P4b} \\
		&\qquad\qquad \left \Vert \begin{matrix}
			\phi_{H,l}^2 N {\rm{vec}}(\bold{\Phi}_k) \\
			\sqrt{2N}\phi_{H,l}\bold{\Phi}_k (\boldsymbol{h}_l+\overline{\bold{H}}_l^H\boldsymbol{v})
		\end{matrix} \right \Vert  \leq x_{l,k},  \label{P4c} \\
		&\qquad\qquad  
		y_{l,k}\bold{I}+\phi_{H,l}^2N\bold{\Phi}_k \succeq 0, \enspace y_{l,k} \geq 0, \label{P4d} \\
		&\qquad\qquad  
		\bold{W}_k \succeq 0, \enspace {\rm{rank}}(\bold{W}_k)=1, \label{P4e} \\
		& \qquad\qquad {\rm{Tr}}\{\bold{W}_{j+1}\}\! \leq\! {\rm{Tr}}\{\bold{W}_{j}\},\forall k \!\in\! \mathcal{K}, \forall j \!\in\! \mathcal{S}.
	\end{align}
\end{subequations}

\Needspace{10\baselineskip}

According to \cite{ref36}, the constraint~\eqref{P4e} is non-convex due to the existing rank-one constraint. Therefore, the rank-one constraint is discarded using the SDR technique, and then the problem is solved using the convex optimization tool CVX \cite{ref39,ref40}. Although the optimal solution obtained in most cases of solving this problem is a rank-one solution, there are still cases where a rank-one solution is unavailable. Therefore, we use a Gaussian randomization technique to reconstruct the rank-one solution from the optimal solution to obtain an efficient solution to the optimization problem \cite{ref39,ref40}.

\subsubsection{Reflecting Phase Optimization}
$\bold{P3}$ is to find feasible solution $\boldsymbol{v}^*$ of $\boldsymbol{v}$ at a fixed $\{\boldsymbol{w}_k\}_{k=1}^K$ and is solved by the SDR technique. Then, $c_{l,k}$ is rewritten as
\begin{equation}
	\label{eq:pcu-clk}
	\begin{split}
		c_{l,k}^e &= {\rm{Tr}}\{\bold{C}_{l,k}\overline{\bold{V}}\} + \boldsymbol{h}_l^H \bold{\Phi}_k \boldsymbol{h}_l-(1+\kappa_r)\sigma_k^2,
	\end{split}
\end{equation}
where $\overline{\bold{V}}=\overline{\boldsymbol{v}}\overline{\boldsymbol{v}}^H$, $\overline{\boldsymbol{v}}=[\boldsymbol{v}^H \enspace 1]^H$, $\bold{C}_{l,k} = 
\begin{bmatrix}
	\overline{\bold{H}}_l \bold{\Phi}_k \overline{\bold{H}}_l^H & \overline{\bold{H}}_l \bold{\Phi}_k \boldsymbol{h}_l \\
	\boldsymbol{h}_l^H \bold{\Phi}_k \overline{\bold{H}}_l^H & 0
\end{bmatrix}, \enspace \forall k\in \mathcal{K}, \forall l\in \mathcal{T}$.

Similarly, the second equation in~\eqref{eq:pcu-components} can be rewritten as
\begin{equation}
	\label{eq:pcu-norm-rewrite}
	\begin{split}
		\Vert (\boldsymbol{v}^H \overline{\bold{H}}_l + \boldsymbol{h}_l^H) \bold{\Phi}_k \Vert_2^2 = {\rm{Tr}} \{ \bold{D}_{l,k}\overline{\bold{V}} \} + \boldsymbol{h}_l^H \bold{\Phi}_k \bold{\Phi}_k \boldsymbol{h}_l,
	\end{split}
\end{equation}
where $\bold{D}_{l,k} = 
\begin{bmatrix}
	\overline{\bold{H}}_l \bold{\Phi}_k \bold{\Phi}_k^H \overline{\bold{H}}_l^H & \overline{\bold{H}}_l \bold{\Phi}_k\bold{\Phi}_k^H \boldsymbol{h}_l \\
	\boldsymbol{h}_l^H \bold{\Phi}_k \bold{\Phi}_k^H \overline{\bold{H}}_l^H & 0
\end{bmatrix}$.

With variable $\overline{\bold{V}}$, \eqref{P4c} is reformulated using SCA \cite{ref35}.
\begin{equation}
	\label{eq:pcu-sca}
	\begin{split}
		\phi_{H,l}^4 N^2 \Vert \bold{\Phi}_k \Vert_F^2\! +\!2\phi_{H,l}^2 N \big({\rm{Tr}} \{ \bold{D}_{l,k}\overline{\bold{V}} \} \!+\! \boldsymbol{h}_l^H \bold{\Phi}_k \bold{\Phi}_k \boldsymbol{h}_l \big) \\ \leq 2Re\{ x_{l,k}^{e,(n)}x_{l,k}^e \} - \vert x_{l,k}^{e,(n)} \vert^2, 
	\end{split}
\end{equation}
where $x_{l,k}^{e,(n)}$ is optimal solution obtained at the $n$th iteration.

Moreover, the constraint~\eqref{P4d} is independent of $\boldsymbol{v}$. According to \textbf{Lemma 1}, we can obtain $y_{l,k}^e = \lambda^+(\bold{Q}_{l,k})$. Based on matrix properties \cite{ref37,ref38}, we can obtain that $\lambda(\bold{Q}_{l,k})=\phi_{H,l}^2 N \lambda(\bold{\Phi}_k), \enspace \forall k\in \mathcal{K}, l \in \mathcal{T}$.

Therefore, $\bold{P3}$ can be transformed into $\bold{P5}$.
\vspace{-6pt}
\begin{subequations}
	\label{P5}
	\begin{align}
		\bold{P5}:	&\mathop{{\rm{Find}}} \quad 
		\overline{\bold{V}} \label{P5a} \\
		&\enspace \enspace s.t. \quad\enspace
		\phi_{H,l}^2 N {\rm{Tr}}\{\bold{\Phi}_k\} - \sqrt{2 {\rm{log}}\,(1/P_{out}^k)}x_{l,k}^e \nonumber \\  &\qquad\qquad\qquad\qquad +{\rm{log}}\,(P_{out}^k)y_{l,k}^e +c_{l,k}^e\geq 0, \label{P5b} \\
		&\qquad\qquad  \eqref{eq:pcu-sca}, \enspace
		\overline{\bold{V}} \succeq 0, \enspace \widetilde{{\rm{diag}}}(\overline{\bold{V}})=\bold{I}_{N+1}, \label{P5c} \\ 
		&\qquad\qquad {\rm{rank}}(\overline{\bold{V}})=1, \enspace \forall k\in \mathcal{K}, \forall l\in \mathcal{T}. \label{P5d}
	\end{align}
	\vspace{-16pt}
\end{subequations}

$\bold{P4}$ is the objective function minimization problem, and $\bold{P5}$ is the problem of finding feasible solutions under constraints. The AO of $\bold{P4}$ and $\bold{P5}$ obtains the optimal solution to the original problem. However, the feasible set of $\bold{P5}$ becomes smaller because $\bold{P4}$ minimizes $\{\boldsymbol{w}_k\}_{k=1}^K$, and it is difficult to find the optimal value. Inspired by \cite{ref21}, we introduce the relaxation variable $\boldsymbol{\alpha}$ with $\{\widetilde{{\rm{diag}}}(\widehat{\bold{V}})\preceq {\rm{diag}(\boldsymbol{\alpha})},\enspace \boldsymbol{\alpha} = [\alpha_1,\alpha_2,\dots,\alpha_{N+1}], \enspace \alpha_n > 0, \enspace \forall n \in \mathcal{N} = \{ 1,2,\dots,N+1 \}$ instead of $\widetilde{{\rm{diag}}}(\overline{\bold{V}})=\bold{I}_{N+1}$, named $\boldsymbol{\alpha}$-relaxation. The feasible set is compressed or expanded by the size change of $\boldsymbol{\alpha}$ to find the feasible solution efficiently. In addition, it is necessary to constrain the change of $\boldsymbol{\alpha}$ by $\mathop{{\rm{min}}} \boldsymbol{\alpha}$ to prevent the continuous expansion of the feasible domain. Thus, $\bold{P5}$ is converted to
\vspace{-6pt}
\begin{subequations}
	\label{P6}
	\begin{align}
		\bold{P6}:	&\mathop{{\rm{min}}} \limits_{\widehat{\mathbf{V}}, \boldsymbol{x}^e, \boldsymbol{y}^e,\boldsymbol{\alpha}} \quad 
		\sum_{n=1}^{N+1} \alpha_n \label{P6a} \\
		&\enspace \enspace \text{s.t.} \quad
			\eqref{eq:pcu-sca},  \eqref{P5b},  
			\widehat{\mathbf{V}} \succeq 0,  \widetilde{{\rm{diag}}}(\widehat{\mathbf{V}})\preceq {\rm{diag}}\left(\boldsymbol{\alpha}\right), \label{P6b} \\
		&\ \ \ \ \qquad {\rm{rank}}(\widehat{\mathbf{V}})=1,   \alpha_n>0,\forall n \in \mathcal{N}. \label{P6c}
	\end{align}
	\vspace{-16pt}
\end{subequations}

Similarly, SDR is used to relax the rank-one constraint and the backtracking method is used to obtain better solution \cite{10057422,ref25}.  Meanwhile, by the above $\alpha$-relaxation, the optimal solution $\big(\widehat{\bold{V}}^o,\{\boldsymbol{w}_k\}_{k=1}^K\big)$ can be obtained. \textbf{P6} is a relaxed problem of \textbf{P5} by relaxing the constraint $\widetilde{{\rm{diag}}}(\overline{\bold{V}})=\bold{I}_{N+1}$ for $\widetilde{{\rm{diag}}}(\widehat{\bold{V}})\preceq {\rm{diag}(\boldsymbol{\alpha})}$, which is more simple to obtain a feasible solution. However, the feasible solution $\widehat{\bold{V}}^o$ obtained for problem \textbf{P6} does not necessarily satisfy the constraint $\{ \widetilde{{\rm{diag}}}(\widehat{\bold{V}})=\bold{I}_{N+1}, \, {\rm{rank}}(\widehat{\bold{V}})=1\}$, where $\{ \widetilde{{\rm{diag}}}(\widehat{\bold{V}})=\bold{I}_{N+1}, \, {\rm{rank}}(\widehat{\bold{V}})=1\}$ is fixed-rank ellipses that is difficult to be solved. In order to satisfy the rank-one constraint, we use the Gaussian randomization technique to reconstruct rank-one solution $\widehat{\boldsymbol{v}}^o$ \cite{ref39,ref40}. In order to satisfy the constraint $ \vert \widehat{v}_n^o \vert=1, \, \forall n \in \mathcal{N} $, we use the projection operation $ \boldsymbol{\overline{v}}^o = \Pi_{\mathcal{C}} (\widehat{\boldsymbol{v}}^o)$ to project $\widehat{\boldsymbol{v}}^o$ into the constraint space of \textbf{P5}, where $\Pi_{\mathcal{C}} (\widehat{\boldsymbol{v}}^o)$ is the element-wise projection of $\widehat{\boldsymbol{v}}^o$ as follows $\Pi_{\mathcal{C}} (\widehat{\boldsymbol{v}}^o):= \mathop{{\rm{argmin}}}\limits_{\boldsymbol{y} \in \mathcal{C}}  \Vert \widehat{\boldsymbol{v}}^o - \boldsymbol{y} \Vert_2, \, \mathcal{C} = \{ \vert y_n \vert=1, \, \forall n \in \mathcal{N} \} $.
\vspace{-8pt}
\begin{equation}
	\label{eq:projection}
	\Pi_{\mathcal{C}} (\widehat{\boldsymbol{v}}^o) = \left\{
	\begin{aligned}
		&\widehat{v}_n^o, \enspace\quad \text{if} \enspace 
		\vert \widehat{v}_n^o \vert = 1, 
		\, \forall n \in \mathcal{N}, \\
		&\frac{\widehat{v}_n^o}{\vert \widehat{v}_n^o \vert}, 
		\enspace\  \text{otherwise},\\
	\end{aligned}
	\right.
	\vspace{-6pt}
\end{equation}
It should be noted that when the SDR solution satisfies the rank-one constraint, the corresponding solution vector can be directly recovered without additional performance loss caused by rank-one recovery. In contrast, when the relaxed SDR solution is not rank one, Gaussian randomization is required for rank-one recovery, followed by element-wise projection to satisfy the unit-modulus constraint, which may introduce a certain optimality gap. Therefore, in the non-rank-one case, the recovered solution is generally a feasible sub-optimal solution to the original non-convex problem \cite{ref39,ref40}.

\vspace{-30pt}
\subsection{FCU Scenario}
\vspace{-6pt}
\textbf{Proposition 3:} Based on Section \uppercase\expandafter{\romannumeral2}-B and matrix properties \cite{ref37,ref38}, the inequality~\eqref{Pr_2} can be converted to
\begin{equation}
	\label{eq:fcu-chance}
	\begin{split}
		{\rm{Pr}}\{ \widetilde{\bold{\Lambda}}_l^H \widetilde{\bold{Q}}_{l,k} \widetilde{\bold{\Lambda}}_l + 2Re \{ \widetilde{\boldsymbol{r}}_{l,k}^H \widetilde{\bold{\Lambda}}_l \} + \widetilde{c}_{l,k} \geq 0 \} \\ \geq 1 - P_{out}^k, \enspace \forall k\in \mathcal{K}, \forall l\in \mathcal{T},
	\end{split}
\end{equation}
where
\vspace{-0.2cm}
\begin{equation}
	\label{eq:fcu-definitions}
	\begin{split}
		\widetilde{\boldsymbol{r}}_{l,k} &= 
		\begin{bmatrix}
			\phi_{h,l} \bold{\Phi}_k(\overline{\boldsymbol{h}}_l+\overline{\bold{H}}_l^H \boldsymbol{v}) \\
			\phi_{H,l} {\rm{vec}}^*(\boldsymbol{v}(\boldsymbol{v}^H\overline{\bold{H}}_l+\overline{\boldsymbol{h}}^H_l)\bold{\Phi}_k)
		\end{bmatrix},\\
		\widetilde{\bold{Q}}_{l,k} &= 
		\begin{bmatrix}
			\phi_{h,l}^2 \bold{\Phi}_k & \phi_{h,l}\phi_{H,l}(\bold{\Phi}_k\otimes\boldsymbol{v}^T) \\
			\phi_{h,l}\phi_{H,l}(\bold{\Phi}_k\otimes\boldsymbol{v}^*) & \phi_{H,l}^2 (\bold{\Phi}_k \otimes \bold{V}^T)
		\end{bmatrix}, \\
		\widetilde{\bold{\Lambda}}_l &= [\bold{\Lambda}_{h,l}^H \enspace \bold{\Lambda}_{H,l}^T]^H, \enspace 	\widetilde{c}_{l,k} = {c}_{l,k}, \enspace \forall k\in \mathcal{K},\forall l\in \mathcal{T}.
	\end{split}
\end{equation}

{\it{Proof:}} See Appendix D.	\hfill$\blacksquare$

Similarly, we can obtain
\begin{equation}
	\label{eq:fcu-components}
	\begin{split}
		{\rm{Tr}}\{ \widetilde{\bold{Q}}_{l,k}\} &= (\phi_{h,l}^2 + \phi_{H,l}^2 N){\rm{Tr}}\{ \bold{\Phi}_k \}, \\
		\Vert \widetilde{\boldsymbol{r}}_{l,k} \Vert_2^2 &= (\phi_{h,l}^2 + \phi_{H,l}^2 N)\Vert (\boldsymbol{v}^H\overline{\bold{H}}_l + \overline{\boldsymbol{h}}_l^H)\bold{\Phi}_k \Vert_2^2, \\
		\Vert \widetilde{\bold{Q}}_{l,k} \Vert_F^2 &= (\phi_{h,l}^2 + \phi_{H,l}^2 N)^2 \Vert \bold{\Phi}_k \Vert_F^2,  \forall k\in \mathcal{K},\forall l\in \mathcal{T}.
	\end{split}
\end{equation}

\subsubsection{Active Beamforming Optimization}
We introduce the auxiliary variables $\widetilde{\boldsymbol{x}} \!=\! [\widetilde{x}_{1,1},\widetilde{x}_{2,1},...,\widetilde{x}_{l,k},...,\widetilde{x}_{K,K}]^T$, $\widetilde{\boldsymbol{y}} \!=\! [\widetilde{y}_{1,1},...,\widetilde{y}_{l,k},...,\widetilde{y}_{K,K}]^T$. Applying \textbf{Lemma 1}, the outage probability constraint~\eqref{eq:fcu-chance} can be converted into the deterministic form as 
\begin{equation}
	\label{eq:fcu-deterministic}
	\begin{split}
		&(\phi_{h,l}^2 + \phi_{H,l}^2 N){\rm{Tr}}\{ \bold{\Phi}_k \} - \sqrt{ 2{\rm{log}}\,(1/P_{out}^k) } \widetilde{x}_{l,k}  \\  & \qquad\qquad\qquad\qquad\qquad + {\rm{log}}\,(P_{out}^k)\widetilde{y}_{l,k} + \widetilde{c}_{l,k} \geq 0,  \\
		&\left \Vert \begin{matrix}
			(\phi_{h,l}^2+\phi_{H,l}^2 N) {\rm{vec}}(\bold{\Phi}_k) \\
			\sqrt{2(\phi_{h,l}^2+\phi_{H,l}^2 N)}\bold{\Phi}_k (\widetilde{\boldsymbol{h}_l}+\overline{\bold{H}}_l^H\boldsymbol{v})
		\end{matrix} \right \Vert  \leq \widetilde{x}_{l,k}, \\
		&\widetilde{y}_{l,k}\bold{I} + (\phi_{h,l}^2 + \phi_{H,l}^2 N)\bold{\Phi}_k \succeq 0, \enspace \widetilde{y}_{l,k} \geq 0.
	\end{split}
\end{equation}

Therefore, the problem $\bold{P2}$ can be converted to
\begin{subequations}
	\label{P7}
	\begin{align}
		\bold{P7}:	&\mathop{{\rm{min}}}\limits_{\{\bold{W}_k\}_{k=1}^K,\widetilde{\boldsymbol{u}},\widetilde{\boldsymbol{v}}} \quad 
		\sum_{k=1}^{K}{\rm{Tr}}\{\bold{W}_k\} \label{P7a} \\
		& \enspace s.t. \quad  \eqref{eq:fcu-deterministic}, \enspace \bold{W}_k \succeq 0, \enspace {\rm{rank}}(\bold{W}_k)=1, \forall k\in \mathcal{K},  \label{P7b} \\
		&\qquad\quad
		{\rm{Tr}}\{\bold{W}_{j+1}\} \leq {\rm{Tr}}\{\bold{W}_{j}\}, \enspace \forall j\in \mathcal{S}. \label{P7c}
	\end{align}
\end{subequations}

$\bold{P7}$ is solved using the same method as that used to solve $\bold{P4}$ in the PCU scenario.

\subsubsection{Reflecting Phase Optimization}
As with the reflection phase optimization in the PCU scenario, \eqref{eq:fcu-deterministic} can be rewritten as \vspace{-0.3cm}
\begin{equation}
	\label{eq:fcu-phase-rewrite}
	\begin{split}
		&\varphi{\rm{Tr}}\{ \bold{\Phi}_k \} \!-\! \sqrt{ 2{\rm{log}}\,(1/P_{out}^k) } \widetilde{x}_{l,k}^e \!\! {\rm{log}}\,(P_{out}^k)\widetilde{y}_{l,k}^e \!+\! \widetilde{c}_{l,k}^e \!\geq\! 0,  \\
		&\varphi\Vert \bold{\Phi}_k \Vert_F^2 + 2\varphi({\rm{Tr}} \{ \widetilde{\bold{D}}_{l,k} \check{\bold{V}} \} + \overline{\boldsymbol{h}}_l^H \bold{\Phi}_k \bold{\Phi}_k^H \overline{\boldsymbol{h}}_l) \\
		&\qquad\qquad\qquad\qquad \leq 2Re\{ \widetilde{x}_{l,k}^{e,(n)} \widetilde{x}_{l,k} \} - \vert \widetilde{x}_{l,k}^{e,(n)} \vert^2,
	\end{split}
\end{equation}
where $\varphi = (\phi_{h,l}^2 + \phi_{H,l}^2 N)$, $\widetilde{c}_{l,k}^e = {\rm{Tr}}\{ \widetilde{\bold{C}}_{l,k} \check{\bold{V}} \} + \overline{\boldsymbol{h}}_l^H \bold{\Phi}_k \overline{\boldsymbol{h}}_l - (1+\kappa_r)\sigma_k^2$, $\widetilde{\bold{C}}_{l,k} = 
\begin{bmatrix}
	\overline{\bold{H}}_l \bold{\Phi}_k \overline{\bold{H}}_l^H & \overline{\bold{H}}_l \bold{\Phi}_k \overline{\boldsymbol{h}}_l \\
	\overline{\boldsymbol{h}}_l^H \bold{\Phi}_k \overline{\bold{H}}_l^H & 0
\end{bmatrix}$, and $\widetilde{\bold{D}}_{l,k} = 
\begin{bmatrix}
	\overline{\bold{H}}_l \bold{\Phi}_k\bold{\Phi}_k^H \overline{\bold{H}}_l^H & \overline{\bold{H}}_l \bold{\Phi}_k \bold{\Phi}_k^H \overline{\boldsymbol{h}}_l \\
	\overline{\boldsymbol{h}}_l^H \bold{\Phi}_k \bold{\Phi}_k^H \overline{\bold{H}}_l^H & 0
\end{bmatrix},\enspace \forall k\in \mathcal{K}, \forall l\in \mathcal{T}$.

The same derivation as in $\bold{P6}$, introducing the auxiliary variable $\widetilde{\boldsymbol{\alpha}} = [\widetilde{\alpha}_1,\widetilde{\alpha}_2,\dots,\widetilde{\alpha}_{N+1}]$. $\bold{P3}$ is approximated as
\begin{subequations}
	\label{P8}
	\begin{align}
		\bold{P8}:	&\mathop{{\rm{min}}} \limits_{\check{\bold{V}}, \widetilde{\boldsymbol{x}}^e, \widetilde{\boldsymbol{y}}^e, \widetilde{\boldsymbol{\alpha}} } \quad 
		\sum_{n=1}^{N+1} \widetilde{\alpha}_n \label{P8a} \\
		&\enspace\enspace\enspace  s.t. \quad  \eqref{eq:fcu-phase-rewrite}, \enspace \check{\bold{V}} \succeq 0, \enspace \widetilde{{\rm{diag}}}(\check{\bold{V}}) \preceq {\rm{diag}(\widetilde{\boldsymbol{\alpha}})}, \label{P8b} \\
		&\qquad\qquad {\rm{rank}}(\check{\bold{V}})=1, \enspace \widetilde{\alpha}_n>0,\forall n \in \mathcal{N}. \label{P8c}
	\end{align}
\end{subequations}

$\bold{P8}$ is solved using the same method as that used to solve $\bold{P6}$ in the PCU scenario.

\section{Alternating Optimization Algorithm and Complexity Analysis}
\subsection{Alternating Optimization Algorithm}

Problem $\bold{P1}$ is converted into active and passive beamforming optimization in the above sections. \textbf{Algorithm 1} solves the primal problem by using AO algorithm. 
\begin{algorithm}[H]
	\caption{Alternating Optimization Algorithm for $\bold{P1}$}\label{alg:alg1}
	\begin{algorithmic}
		\STATE 
		\STATE \begin{itemize}
			\item[1:]
			$\textbf{Initialization:}$ Randomly initialize transmit beamforming vectors $\{ \boldsymbol{w}_k^{(0)}\}_{k=1}^K$ and the reflection phase vector $\boldsymbol{v}^{(0)}$. Set the number of iteration $n=1$.
		\end{itemize}
		\STATE \begin{itemize}
			\item[2:]
			$\textbf{Repeat}$
		\end{itemize}
		\STATE \begin{itemize}
			\item[3:]
			According to $\boldsymbol{v}^{(n)}$, $\{\boldsymbol{w}_k^{(n)}\}_{k=1}^K$, solve $\bold{P4}$ or $\bold{P7}$ through CVX and obtain the optimal solution $\{ \boldsymbol{w}_k^{(n+1)}\}_{k=1}^K$.
		\end{itemize}
		\STATE \begin{itemize}
			\item[4:]
			According to $\boldsymbol{v}^{(n)}$, $\{\boldsymbol{w}_k^{(n+1)}\}_{k=1}^K$, solve $\bold{P6}$ or $\bold{P8}$ through CVX and obtain the optimal solution $\boldsymbol{v}^{(n+1)}$.
		\end{itemize}
		\STATE \begin{itemize}
			\item[5:]
			$n=n+1$.
		\end{itemize}
		\STATE \begin{itemize}
			\item[6:]
			\textbf{Until} $\vert \sum_{k=1}^{K} \big( {\rm{Tr}}(\boldsymbol{w}_k^{(n+1)}\boldsymbol{w}_k^{(n+1),H}) - {\rm{Tr}}(\boldsymbol{w}_k^{(n)}\boldsymbol{w}_k^{(n),H}) \big)\vert \leq 10^{-4}$ converges.
		\end{itemize}
		\STATE \begin{itemize}
			\item[7:]
			\textbf{Return} $\boldsymbol{v}^{(n+1)}$ and $\{ \boldsymbol{w}_k^{(n+1)} \}_{k=1}^K$.
		\end{itemize}
	\end{algorithmic}
	\label{alg1}
\end{algorithm}

\subsection{Complexity Analysis}
We consider all constraints to be active constraints, which is a worst-case scenario. These constraint types are second-order cone (SOC) constraints, linear constraints, and linear matrix inequality (LMI) constraints, where the linear constraints complexity is neglected. $\bold{P4}$ and $\bold{P7}$ have the same computational complexity because they differ only in the constraint coefficients. Similarly, $\bold{P6}$ and $\bold{P8}$ have the same computational complexity. 

According to \cite{ref18}, the computational complexity of active and passive beamforming optimization are $\mathcal{O}_{\boldsymbol{w}} = \mathcal{O} \{ [(K^2+K)(M+1)]^{\frac{1}{2}} n_1 [n_1^2 + n_1(K^2+K) M^2+(K^2+K)M^3 +\frac{1}{2}n_1(K^2+K)(M+1)^2M^2 ] \}$ and $\mathcal{O}_{\boldsymbol{v}} = \mathcal{O} \{ [(K^2+K)(N+1)]^{\frac{1}{2}} n_2 [n_2^2 + (n_2+N+1)(K^2+K)(N+1)^2] \}$, where $n_1 = MK$ and $n_2 = N+1$. Therefore, the computational complexity of the AO algorithm is $\mathcal{O}_{AO} = \mathcal{O}( I_{AO} + I_{BT} )(\mathcal{O}_{\boldsymbol{w}} + \mathcal{O}_{\boldsymbol{v}})$, where $I_{AO}$ and $I_{BT}$ indicate the number of AO and backtracking iterations, respectively.

\section{Simulation Results and Discussion}
\begin{figure*}[!t]
	\centering
	\vspace{-0.7cm} 
	
	\subfloat[PCU $\zeta_H=0.01$]{%
		\includegraphics[width=2.8 in]{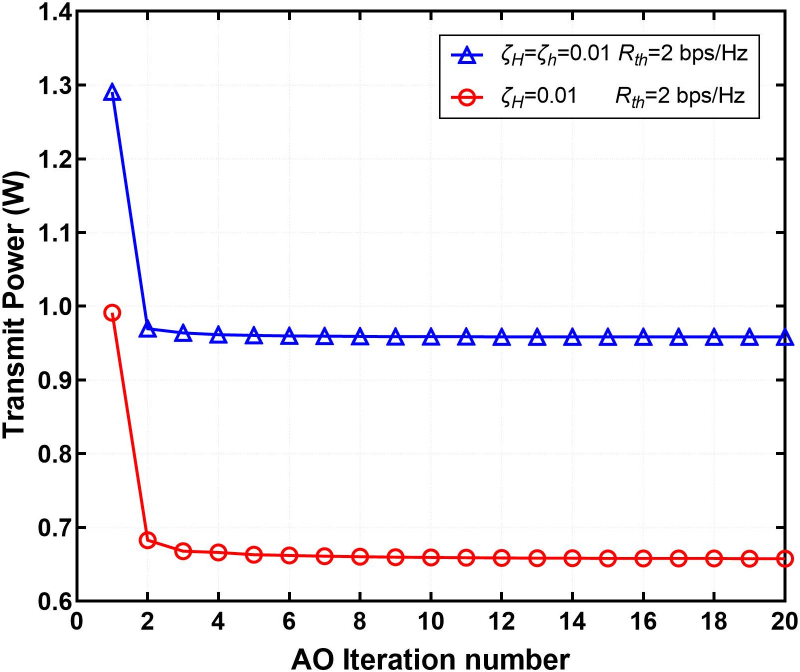}%
		\label{fig02_a}}
	\hfil
	\subfloat[FCU $\zeta_H=\zeta_h=0.01$]{%
		\includegraphics[width=2.8 in]{fig/fig02_a}%
		\label{fig02_b}}
	
	\caption{Our algorithm convergence figure and comparison figure of different algorithms. 
		$M=4$, $N=30$, $K=2$, $\kappa_r=\kappa_t=0.01$, $P_{out}=0.05$.}
	\label{fig02}
	
	\vspace{-0.cm}
\end{figure*}

This section provides relevant numerical results to evaluate our proposed scheme. We assume the BS and RIS are set at (5, 0, 0) and (0, 50, 20) in meters, respectively. NOMA users are uniformly distributed in a circle with (5, 70, 0) in meters as the center and 5m as the radius. Set $\alpha_{BR}=2.2$, $\alpha_{rk}=2$, $\alpha_{bk}=4$, $\beta_0=-30\enspace {\rm{dB}}$, $\sigma_k^2=-80 \enspace {\rm{dBm}}, \forall k\in \mathcal{K}$, $R_{th}^k=R_{th}, k\in \mathcal{K}$, and $P_{out}^k=P_{out}, k\in \mathcal{K}$. The Rician factors are $R_{BR}=R_{rk}=3 \enspace {\rm{dB}}$ \cite{ref35}. 
$M$, $N$, and $K$ are treated as general system parameters, allowing the proposed framework to accommodate different system configurations by adjusting the corresponding system dimensions without redesigning the algorithm.
Moreover, $\phi_{H,k}$ and $\phi_{h,k}$ are defined as $\zeta_H \Vert {\rm{vec}}(\overline{\bold{H}}_k) \Vert_2$ and $\zeta_h \Vert {\rm{vec}}(\overline{\boldsymbol{h}}_k) \Vert_2$, respectively \cite{ref22}. 
For practical NOMA networks, the number of users multiplexed within each NOMA cluster should not be large to reduce SIC complexity and error propagation \cite{9174801, 9264161, 11078432}. Therefore, we consider a maximum of four users within each NOMA cluster.
We use the OMA network and the traditional NOMA network as benchmarks. In addition, `` w/ RIS" and ``w/o RIS" indicate the presence and absence of RIS, respectively. ``w/r" and ``w/n" indicate the robust and non-robust design, respectively.

\begin{figure*}[!t]
	\centering
	\vspace{-0.7cm} 
	\subfloat[PCU $\zeta_H=0.01$]{\includegraphics[width=2.8 in]{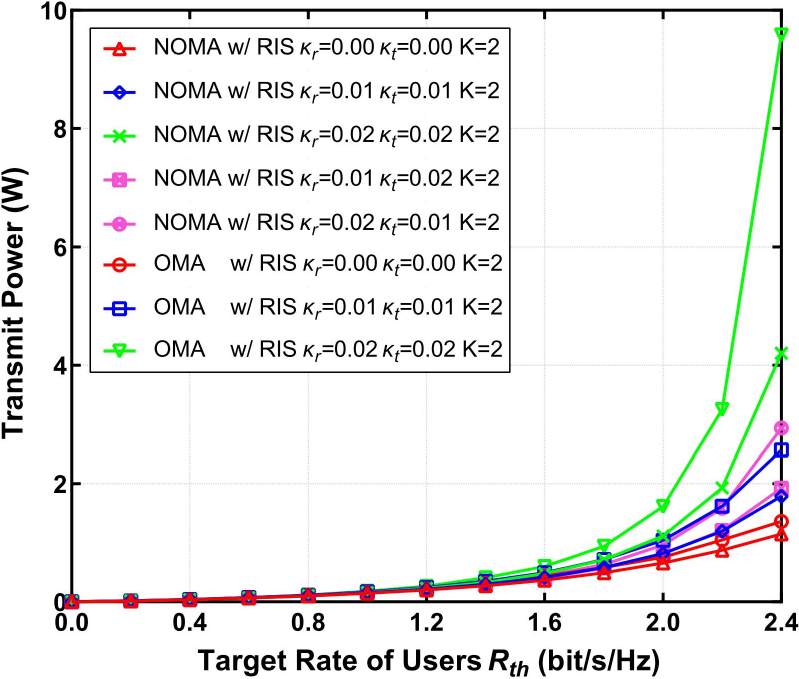}%
		\label{fig03_a}}
	\hfil
	\subfloat[PCU $\zeta_H=0.01$]{\includegraphics[width=2.8 in]{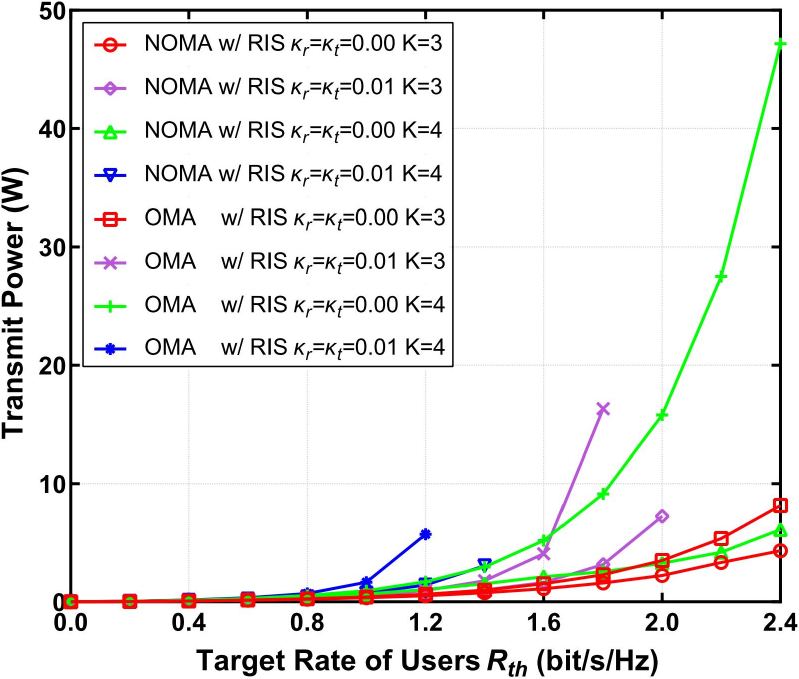}%
		\label{fig03_b}}
	\hfil
	\centering
	\vspace{-0.2cm}
	\subfloat[FCU]{\includegraphics[width=2.8 in]{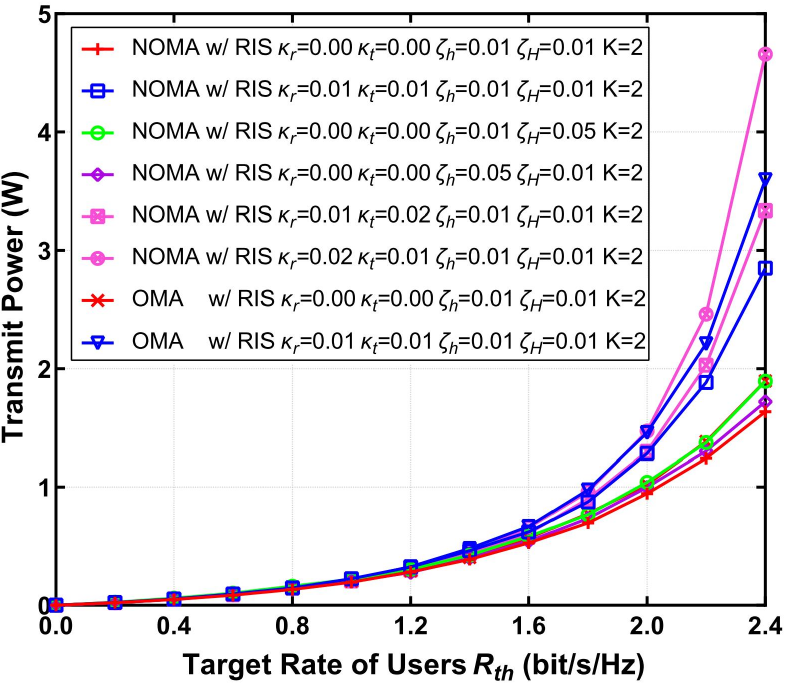}%
		\label{fig03_c}}
	\hfil
	\subfloat[FCU]{\includegraphics[width=2.8 in]{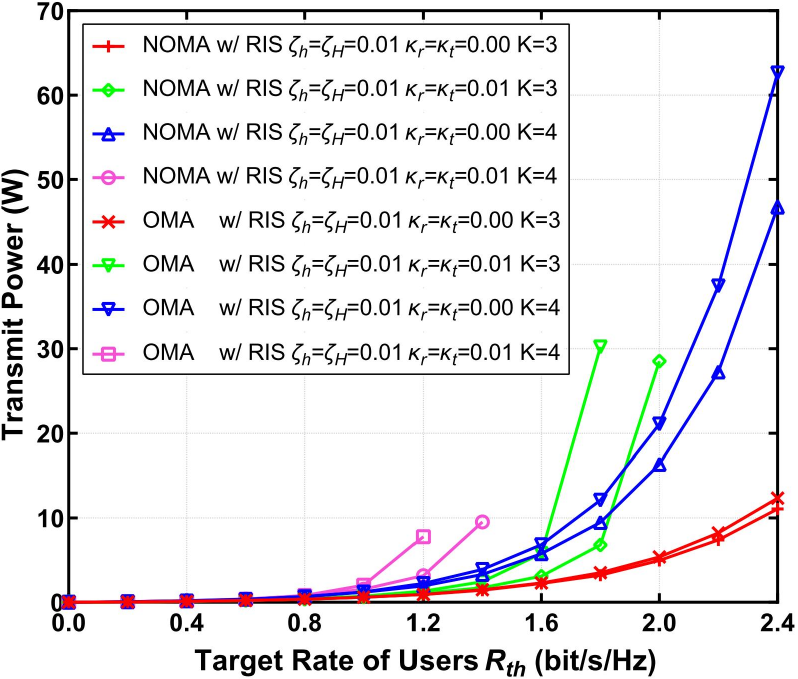}%
		\label{fig03_d}}
	\hfil
	\caption{The minimum BS transmit power with different levels of HWI and different number of users in the RIS-assisted multi-beam NOMA network. $M=4$, $N=30$, $P_{out}=0.05$.}
	\label{fig03}
	\vspace{-0.5cm}
\end{figure*}

As shown in Fig. 2(a), for the case of $M=4$, $N=30$, and $K=2$ under the PCU scenario, the AO algorithm based on the $\alpha$-relaxation is implemented to verify its convergence. It can be observed that the proposed algorithm converges rapidly. 
The rapid convergence enables the proposed framework to update the active beamforming and RIS configuration according to the current CSI, providing a basis for real-time optimization under the considered quasi-static fading model.
In addition, the sphere-bounding method combined with the SDR technique is a representative approach for solving optimization problems subject to outage probability constraints \cite{ref19,ref21}. To further demonstrate the performance advantage of the proposed algorithm, the two algorithms are compared in Fig. 2(b). It can be observed that the proposed BTI-based algorithm requires lower BS transmit power than the sphere-bounding-plus-SDR method. This performance difference is associated with the different treatments of statistical CSI uncertainty adopted by the two methods. Specifically, the sphere-bounding method first confines the random CSI error within a deterministic uncertainty region, whereas the BTI-based method directly exploits the Gaussian statistical structure of the CSI error to construct a convex safe approximation of the outage probability constraint \cite{ref19}. Therefore, the simulation results demonstrate the performance advantage of the proposed optimization framework under the considered simulation settings.

\begin{figure*}[!t]
	\centering
	\vspace{-0.7 cm} 
	\subfloat[PCU $M=4$, $N=30$]{\includegraphics[width=2.8 in]{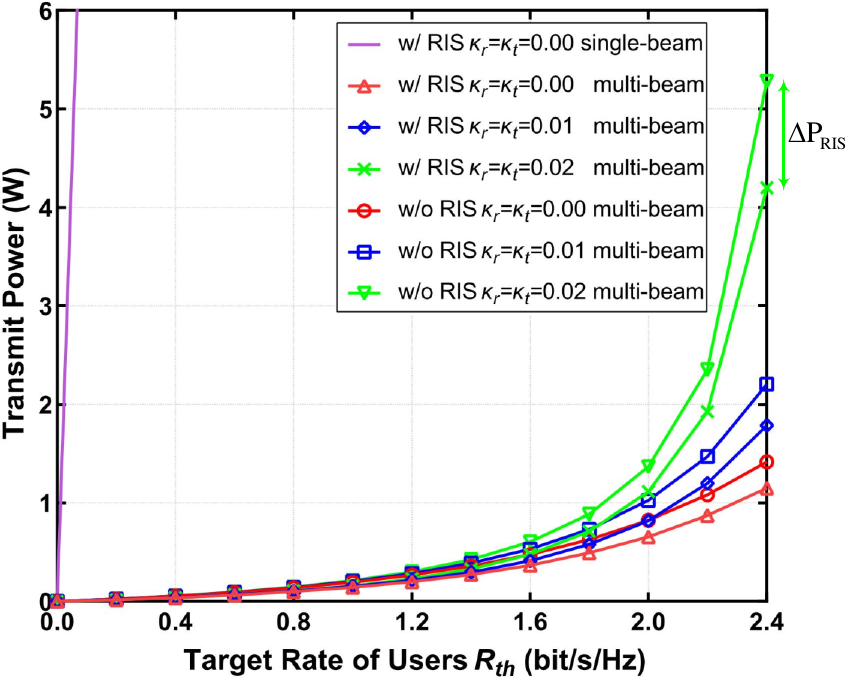}%
		\label{fig04_a}}
	\hfil
	\subfloat[FCU $M=4$, $N=30$]{\includegraphics[width=2.8 in]{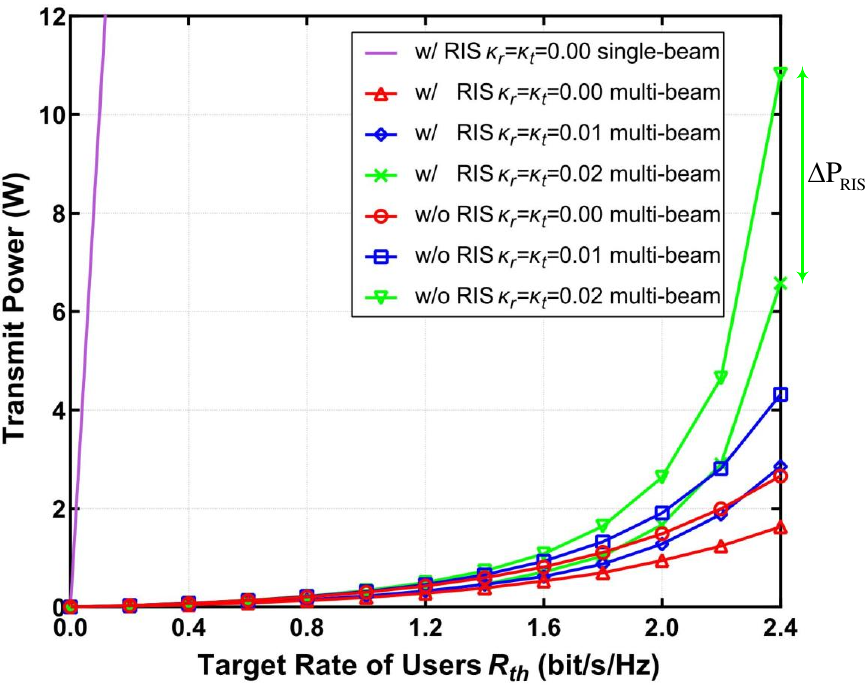}%
		\label{fig04_b}}
	\hfil
	\subfloat[PCU $\kappa_r=\kappa_t=0.00$, $R_{th}=2{\rm{bit/s/Hz}}$]{\includegraphics[width=2.8 in]{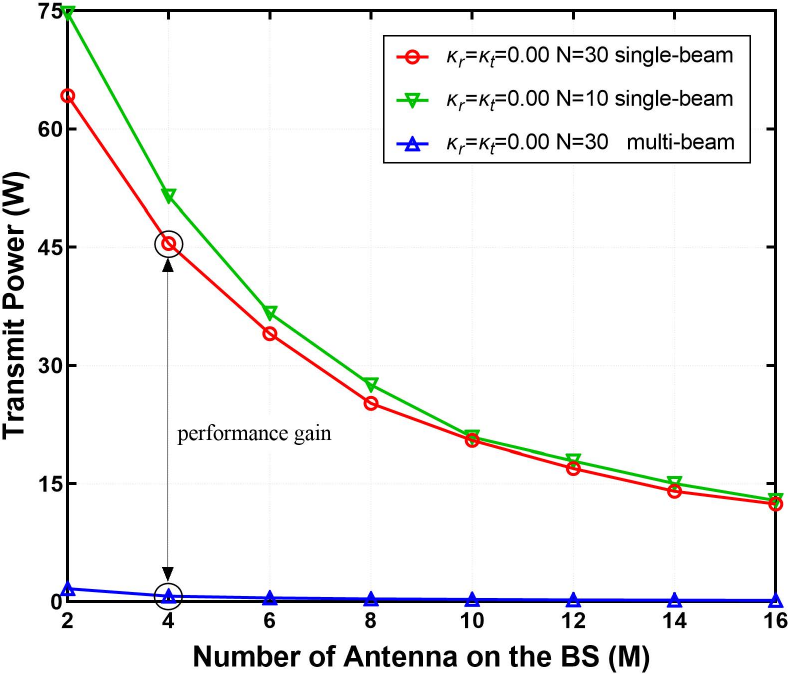}%
		\label{fig04_c}}
	\hfil
	\subfloat[FCU $\kappa_r=\kappa_t=0.00$, $R_{th}=2{\rm{bit/s/Hz}}$]{\includegraphics[width=2.8 in]{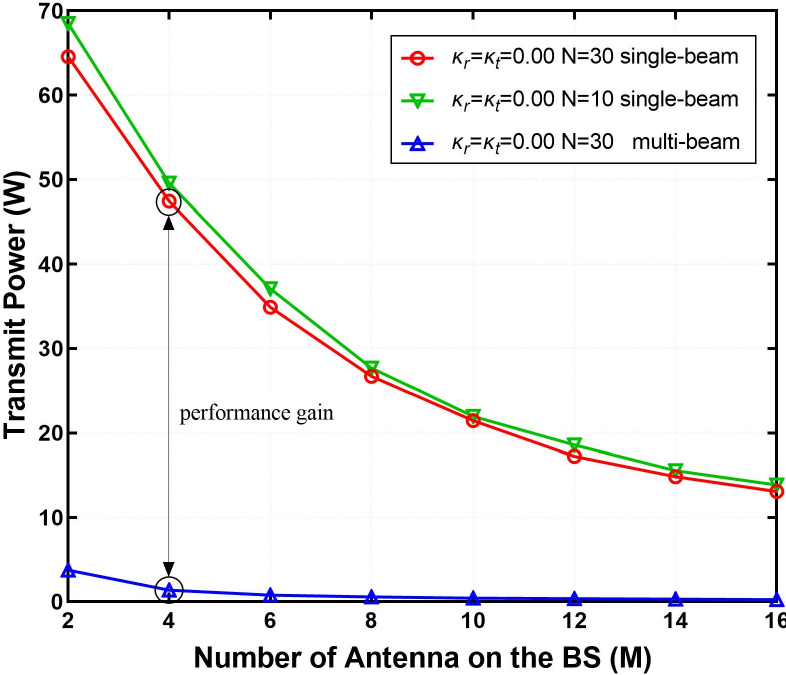}%
		\label{fig04_d}}
	\hfil
	\vspace{-0.2cm} 
	\caption{Performance comparison of RIS-assisted NOMA networks in FCU and PCU scenarios. $K=2$, $P_{out}=0.05$, $\zeta_H=\zeta_h=0.01$.}
	\label{fig04}
	\vspace{-0.5cm}
\end{figure*}

In order to discuss the impact of transceiver HWI and imperfect CSI on network performance, simulation experiments were conducted for two to four user cases under PCU and FCU scenarios, respectively, and the OMA network was used as benchmark for comparison. In Fig. 3, the larger the $\kappa_r$ and $\kappa_t$ in the two-user scenario, the higher the minimum BS transmit power at the same $R_{th}$. In addition, the distortion noise caused by HWI is related to the transmit and receive power, which is a state of negative feedback. 
This phenomenon can be explained by the fact that the HWI-induced distortion noise increases with the transmit and received signal powers, and thus increasing the transmit power cannot fully compensate for the performance degradation caused by severe HWI. Moreover, increasing the number of users introduces more inter-user interference and SIC constraints, which reduces the feasible region and increases the required transmit power.
An increase in the number of users can lead to a smaller feasible set and thus make power matching more difficult or even impossible.
As seen from Fig. 3 (b) and Fig. 3 (d), the higher the number of users, the greater the degree of target rate limitation when HWI is considered. It follows that the number of users is one of the most important factors affecting the achievable rate. Therefore, in a practical communication network, the number of served users should not be large to achieve the desired rate. Moreover, the rate without HWI will be higher than the rate with HWI when the signal is transmitted at the same power. Thus, the $R_{th}$ can not be achieved when not considering HWI non-robust beamforming, which validates the importance of studying robust beamforming. For the same $R_{th}$, NOMA networks always require less transmit power than OMA networks, which reflects the advantage of NOMA networks to improve SE. In addition, in order to obtain higher EE with limited resources, we conducted simulation experiments for the impact of different levels of HWI and CEE on the energy consumption. From Fig. 3 (a) and Fig. 3 (c), it is observed that the receiver HWI and RIS reflection CEE have a greater impact on energy consumption. An increase in the number of users can lead to a smaller feasible set and thus more difficult or even impossible power matching.

\begin{figure*}[!t]
	\centering
	\vspace{-0.7cm}
	\subfloat[PCU $M=4$]{%
		\includegraphics[width=2.8in]{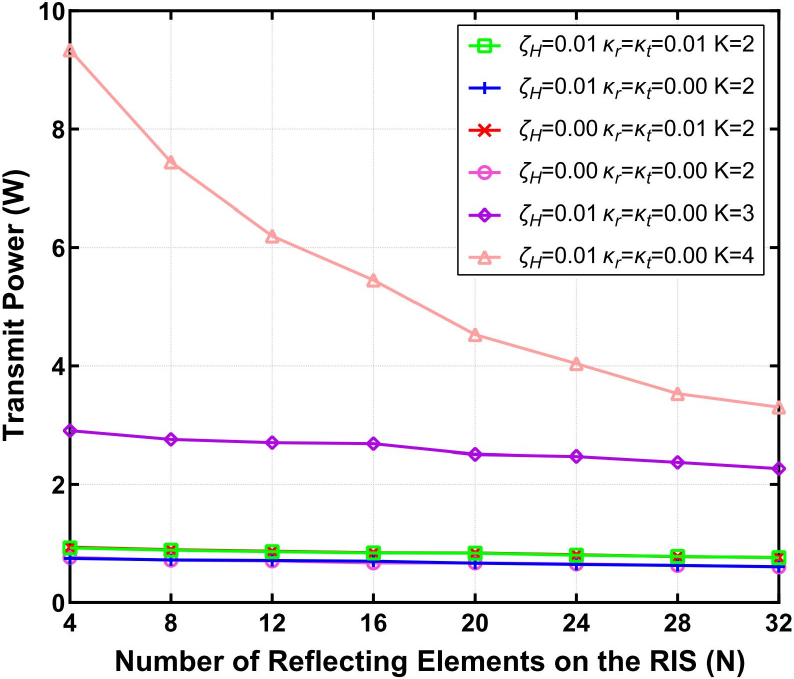}%
		\label{fig05_a}}
	\hfil
	\subfloat[PCU $N=4$]{%
		\includegraphics[width=2.8in]{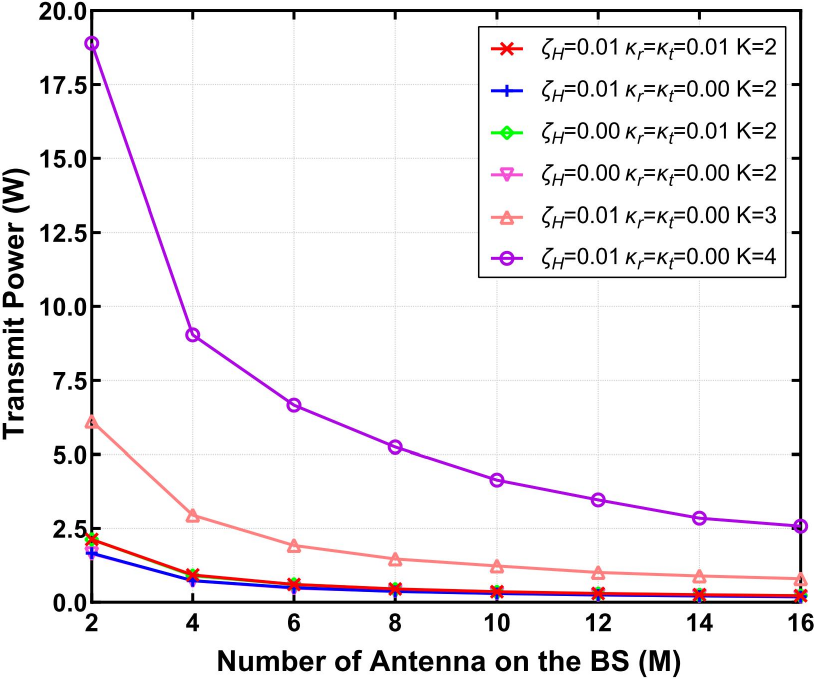}%
		\label{fig05_b}}
	
	\subfloat[FCU $M=4$]{%
		\includegraphics[width=2.8in]{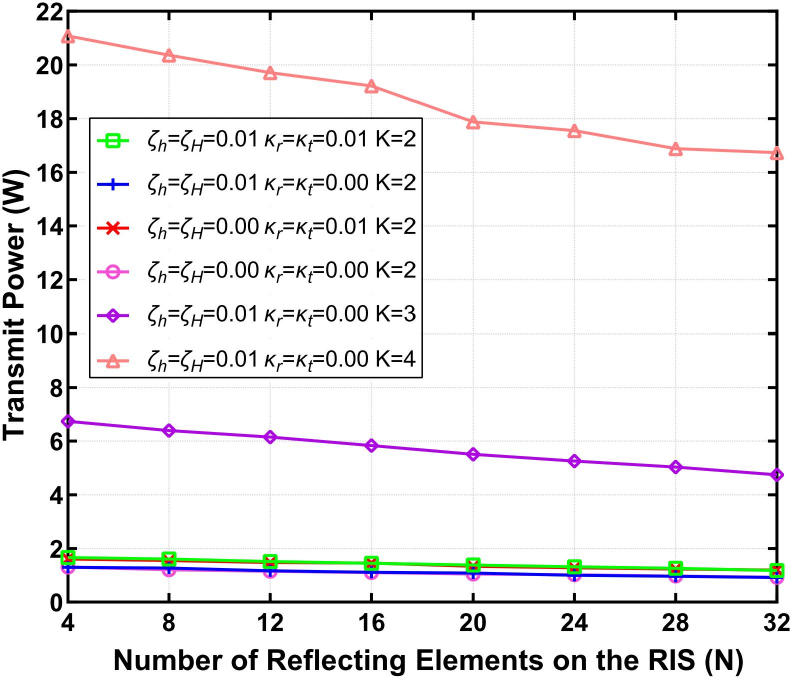}%
		\label{fig05_c}}
	\hfil
	\subfloat[FCU $N=4$]{%
		\includegraphics[width=2.8in]{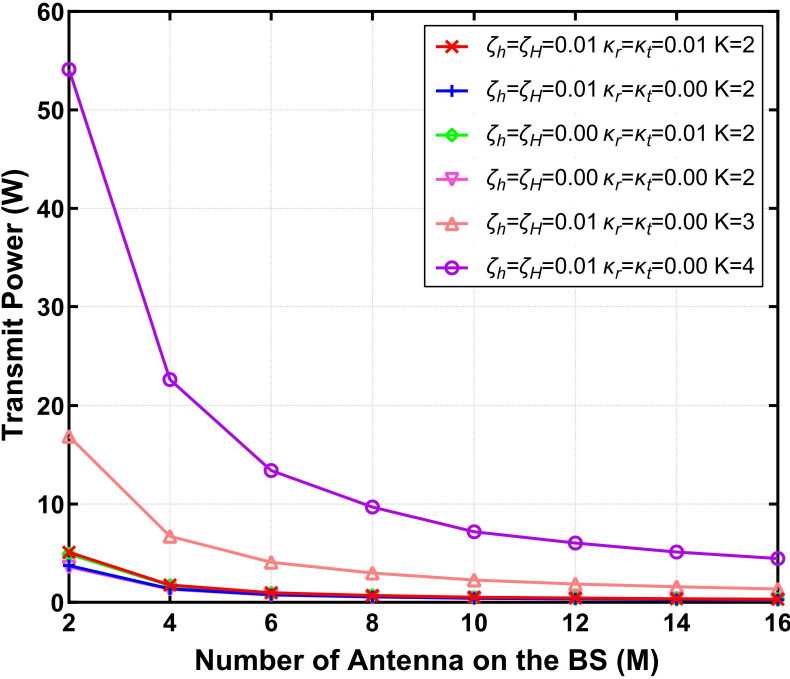}%
		\label{fig05_d}}
	\vspace{-0.2cm}
	\caption{Effect of the number of RIS reflecting elements and BS antennas on the minimum BS transmit power in the RIS-assisted multi-beam NOMA network. $K=2$, $P_{out}=0.05$, and $R_{th}=2~{\rm bit/s/Hz}$.}
	\label{fig05}
	\vspace{-0.cm}
\end{figure*}

Further, we use the traditional NOMA scheme as a benchmark to compare the performance with the RIS-assisted NOMA network to verify the performance improvement with the assistance of RIS. As shown in Fig. 4 (a) and (b), the RIS-assisted NOMA network always requires lower transmit power than the conventional NOMA network when considering the same $R_{th}$ for both PCU and FCU scenarios. 
To make this comparison more explicit, the transmit-power reduction provided by RIS assistance is highlighted by $\Delta P_{\mathrm{RIS}}$ in Fig.~4(a) and Fig.~4(b). Specifically, $\Delta P_{\mathrm{RIS}}=P_{\mathrm{w/o\,RIS}}-P_{\mathrm{w/\,RIS}}$ denotes the transmit-power saving achieved by introducing the RIS under the same HWI level and $R_{th}$.
This power reduction is attributed to the additional reflected link provided by the RIS, which improves the effective channel gain through phase adjustment and consequently reduces the transmit power required to satisfy the same target rate.
Moreover, the BS transmit power is always higher in the FCU scenario than in the PCU scenario. It is shown that the assistance of RIS can effectively reduce energy consumption, while the direct channel imperfect CSI can also cause a further negative impact on the network performance. In addition, from Fig. 4, the lower transmit power is required in the multi-beam NOMA network than in the single-beam NOMA network under the same $R_{th}$. In Fig. 4 (c), the transmit power is 46.63 dBm and 28.63 dBm in the single-beam PCU network and the multi-beam PCU network when $M=4$ and $N=30$, respectively. For this position, the multi-beam network provides a $63\%$ gain. In addition, with the increase of $M$, the performance difference between $N=10$ and $N=30$ becomes smaller, which is due to more compensation of BS antenna gain. 
Similarly, the transmit power is 46.52 dBm and 31.43 dBm in the single-beam FCU network and the multi-beam FCU network when $M=4$ and $N=30$, respectively. For this position, the multi-beam network provides a $48\%$ gain. 
These results confirm the performance advantage of multi-beam NOMA over single-beam transmission. Moreover, user-specific beams provide better spatial selectivity, which is beneficial to physical-layer security. However, such gains are accompanied by higher computational and implementation complexity, since multi-beam transmission requires the joint optimization of multiple user-specific precoders; depending on the transceiver architecture, additional RF-chain resources may also be required. Therefore, practical deployment should balance the transmit-power and potential security benefits against the associated computational and hardware costs.

To investigate the effect of the number of BS antennas and RIS reflecting elements on network performance in a deeper way, we simulated the minimum BS transmit power in these two cases. According to \cite{ref18}, in the case of non-severe imperfect CSI, the transmit power of the BS will decrease as $M$ and $N$ increase. This is because as $M$ and $N$ increase, the degrees of freedom become larger, and the gain in performance increases. However, it can be seen from~\eqref{P4} that an increase in $M$ and $N$ will cause an increase in the impact of imperfect CSI. Therefore, in the actual communication process, the number of BS antennas and RIS reflecting elements need to be limited according to the communication requirements. In addition, according to~\eqref{E_yk_2}, the distortion noise power caused by HWI is directly related to the BS transmit power. Therefore, HWI will have a more serious impact on the network performance. As shown in Fig. 5, small changes in $\zeta_h$ and $\zeta_H$ have a small impact on the BS transmit power, while small changes in HWI will lead to a significant increase in the BS transmit power.

\begin{figure*}[!t]
	\centering
	\subfloat[$M=4$]{%
		\includegraphics[width=0.43\textwidth]{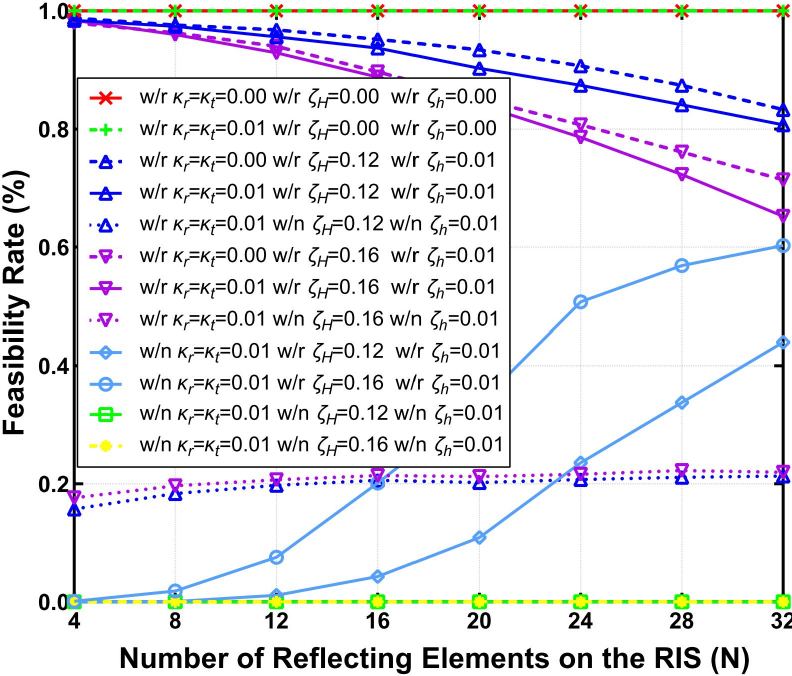}%
		\label{fig06_a}}
	\hspace{0.02\textwidth}
	\subfloat[$N=4$]{%
		\includegraphics[width=0.43\textwidth]{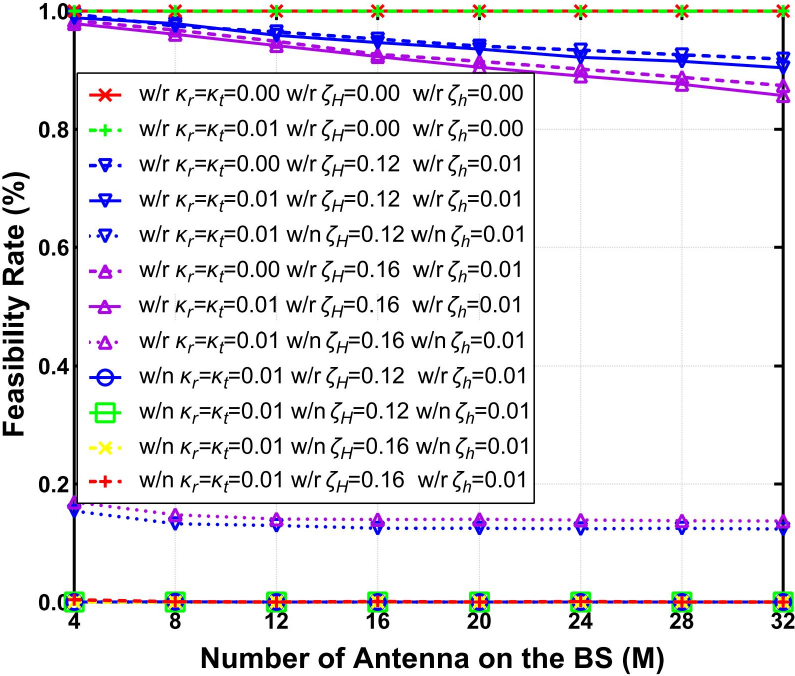}%
		\label{fig06_b}}
	\vspace{-0.2cm}
	\caption{Effect of the number of RIS reflecting elements and BS antennas on the feasibility rate under the multi-beam communication scenario. $P_{out}=0.05$, $R_{th}=2 \enspace {\rm{bit/s/Hz}}$.}
	\label{fig06}
\end{figure*}

\begin{figure*}[!t]
	\centering
	\vspace{-0.3cm}
	\subfloat[PCU]{%
		\includegraphics[width=0.43\textwidth]{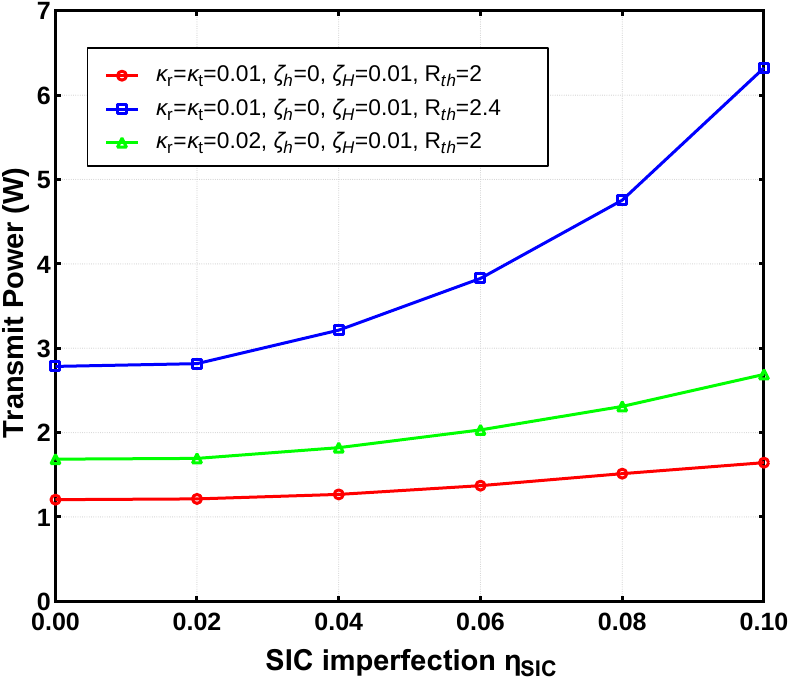}%
		\label{fig07_a}}
	\hspace{0.02\textwidth}
	\subfloat[FCU]{%
		\includegraphics[width=0.43\textwidth]{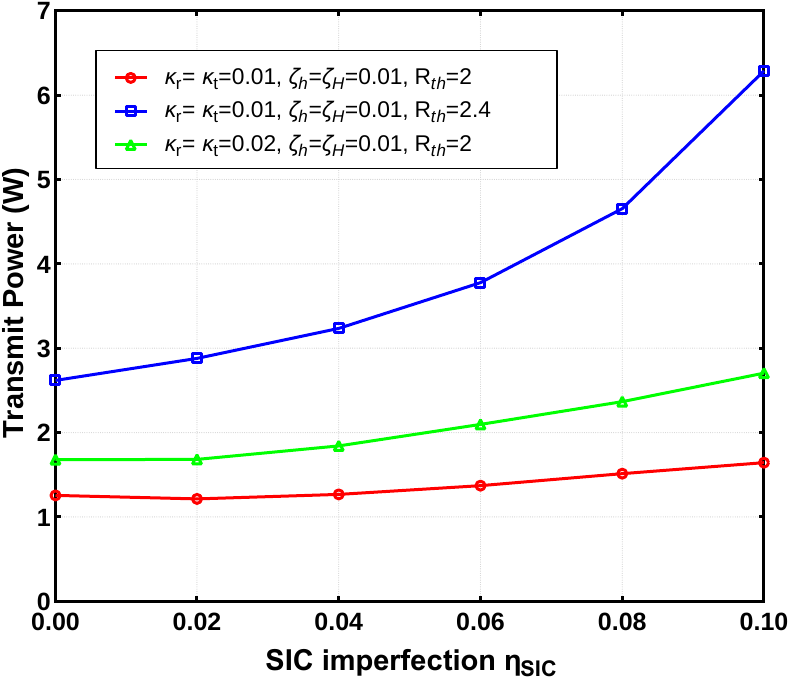}%
		\label{fig07_b}}
	\vspace{-0.2cm}
	\caption{The minimum BS transmit power versus the imperfect SIC factor $\eta_{\mathrm{SIC}}$ under different HWI levels and target rates.}
	\label{fig:imperfect_SIC}
	\vspace{-0.5cm}
\end{figure*}

To specifically describe the extent of the impact of imperfect CSI on the network performance due to the increase of M and N, we conducted simulation experiments on the feasibility rate at different amounts of $M$ and $N$, as shown in Fig.~\ref{fig06}. According to \cite{ref22}, we define the feasibility rate as the ratio of the number of feasible channels to the total number of channels, where the feasible channels denote the channels for which feasible solutions exist that can satisfy the constraint~\eqref{P1b} in $\bold{P1}$. Imperfect CSI is directly related to the BS and RIS, so different amounts of M and N have different effects on network performance. 
As $M$ and $N$ increase, there is a stronger performance gain for communication, but also a concomitant increase in CEE. 
As can be seen in Fig.~\ref{fig06}, the feasibility rate exhibits different trends with increasing $M$ and $N$ under different system settings.
In some cases, this is because the performance gain provided by increasing $M$ or $N$ cannot fully compensate for the effect of imperfect CSI on the network.
Meanwhile, the non-robust beamforming design will be applied in a communication network with imperfect CSI and HWI, which will inevitably result in performance differences. As shown in Fig.~\ref{fig06}, both non-robust HWI and non-robust CSI lead to a decrease in the feasibility rate. 
This comparison further demonstrates the importance of robust beamforming design, since explicitly considering CSI uncertainty and HWI in the optimization process can improve the reliability of satisfying the QoS constraints.
An interesting phenomenon can be seen that the feasibility rate at $\zeta_H=0.16$ in the non-robust scheme is greater than that at $\zeta_H=0.12$, which is since the larger the $\zeta_H=0.16$ the larger the floating range of the channel, which leads to an increase in the feasible probability of the channel under non-robust. In addition, it is known from~\eqref{x_2} and~\eqref{E_yk_2} that the degree of HWI impact increases with increasing BS transmit power. 
Therefore, under non-robust HWI, increasing $N$ can improve the effective channel gain and reduce the required transmit power, thereby alleviating the impact of HWI, while the performance is also affected by imperfect CSI.
Fig.~\ref{fig06}(a) illustrates this effect.
Therefore, factors such as communication scenarios, the number of users, and the required performance increase should be taken into account when selecting $M$ and $N$.
These results indicate that increasing the numbers of BS antennas and RIS reflecting elements does not always guarantee unlimited performance improvement, since the achievable gains are also affected by CSI uncertainty and hardware impairments.
Moreover, as the numbers of BS antennas and RIS reflecting elements increase, the dimension of the optimization variables also expands, which introduces additional challenges for solving the non-convex optimization problem. On the one hand, SDR obtains positive semidefinite matrix solutions by relaxing the rank-one constraints. As the optimization dimension increases, the gap between the relaxed feasible set and the original rank-one constrained feasible set may become larger, thereby affecting the subsequent randomization recovery performance. On the other hand, due to the non-convexity of the original problem, the adopted iterative optimization methods generally can only guarantee convergence to locally optimal solutions or stationary points.
Therefore, the variation in the feasibility rate shown in Fig.~\ref{fig06} is not only affected by CSI uncertainty, but may also be related to the approximation errors in solving high-dimensional optimization problems.

Finally, to further investigate the performance of the proposed scheme under practical SIC conditions, we consider imperfect SIC by introducing a residual SIC factor $\eta_{\mathrm{SIC}}$, where $\eta_{\mathrm{SIC}}=0$ corresponds to perfect SIC and a larger $\eta_{\mathrm{SIC}}$ indicates more severe residual interference after interference cancellation. Specifically, a fraction of the interference from the previously decoded signals remains after SIC. Accordingly, the matrix $\boldsymbol{\Phi}_k$ can be modified as 
$ \widetilde{\boldsymbol{\Phi}}_k
=
\frac{\boldsymbol{w}_k\boldsymbol{w}_k^{H}}
{2^{R_{\mathrm{th}}^k}-1}
-
\sum_{i=k+1}^{K}\boldsymbol{w}_i\boldsymbol{w}_i^{H}
-
\eta_{\mathrm{SIC}}
\sum_{i=1}^{k-1}\boldsymbol{w}_i\boldsymbol{w}_i^{H}
-
\boldsymbol{\Psi},
\quad \forall k\in\mathcal{S},$
and
$	\widetilde{\boldsymbol{\Phi}}_K
=
\frac{\boldsymbol{w}_K\boldsymbol{w}_K^{H}}
{2^{R_{\mathrm{th}}^K}-1}
-
\eta_{\mathrm{SIC}}
\sum_{i=1}^{K-1}\boldsymbol{w}_i\boldsymbol{w}_i^{H}
-
\boldsymbol{\Psi}.$
With this modification, the residual interference caused by imperfect SIC can be incorporated into the same robust optimization framework without changing the overall solution procedure.
Fig.~\ref{fig:imperfect_SIC} shows the BS transmit power versus the imperfect SIC factor $\eta_{\mathrm{SIC}}$ under different HWI levels and target rates. It can be observed that the required BS transmit power generally increases with $\eta_{\mathrm{SIC}}$. This is because more severe residual SIC interference reduces the effective decoding SINR, and thus additional transmit power is required to satisfy the corresponding QoS constraints. Moreover, the required transmit power further increases as the HWI becomes more severe. Similarly, the performance degradation caused by imperfect SIC becomes more pronounced as the target rate increases. This indicates that the system is more sensitive to residual SIC interference under more severe HWI or more stringent QoS requirements. These results demonstrate that residual SIC interference introduces an additional transmit-power penalty compared with the perfect-SIC case. Meanwhile, imperfect SIC can be naturally incorporated into the proposed robust framework through the residual-interference term, further demonstrating the applicability of the proposed scheme under practical SIC conditions.

\section{Conclusion}
In this paper, we investigated robust transmission for RIS-assisted multi-user single-beam and multi-beam NOMA networks under imperfect CSI and transceiver HWI. First, the distortion-noise power caused by HWI was derived and incorporated into the user decoding SINRs, while the CSI uncertainty was characterized through outage-probability constraints. Based on the resulting signal model, the SIC conditions for single-beam and multi-beam transmission were investigated, and a robust transmit-power minimization problem was formulated. Subsequently, an effective algorithm was proposed to solve the resulting non-convex problem. Numerical results verified the effectiveness of the robust design and revealed the impacts of the considered practical impairments on system performance. In particular, as the number of users increases, the limitation imposed by HWI on the achievable rate becomes more pronounced; when the CSI uncertainty is severe, the performance gains brought by increasing the number of BS antennas or RIS reflecting elements are reduced. Moreover, multi-beam NOMA can significantly reduce the required transmit power compared with single-beam transmission.

\appendices
\vspace{-8pt}
\section{Proof of Proposition 1}

From equation~\eqref{y_k}, the undistorted received signal of the $k$th user is
\begin{equation}
	\label{eq:appA-received-signal}
	\hat{y}_k=(\boldsymbol{v}^{H}\bold{H}_k+\boldsymbol{h}_k^{H})\boldsymbol{x}+n_k,\enspace \forall k\in\mathcal{K}.
\end{equation}

Then, we can obtain the variance of distortion noise.
\begin{equation}
	\label{eq:appA-distortion-variance}
	\begin{split}
		\mathbb{E}[|\hat{y}_k|^2]
		&\!=\!\mathbb{E}\left[((\boldsymbol{v}^{H}\bold{H}_k \!+\!\boldsymbol{h}_k^{H})\boldsymbol{x}\!+\! n_k)
		((\boldsymbol{v}^{H}\bold{H}_k \!+\!\boldsymbol{h}_k^{H})\boldsymbol{x}\!+\! n_k)^{H}\right]\\
		&\!=\!(\boldsymbol{v}^{H}\bold{H}_k+\boldsymbol{h}_k^{H})\mathbb{E}[\boldsymbol{x}\boldsymbol{x}^{H}]
		(\boldsymbol{h}_k+\bold{H}_k^{H}\boldsymbol{v})+\sigma_k^2.
	\end{split}
\end{equation}

From equation~\eqref{x_2}, $\mathbb{E}[\boldsymbol{x}\boldsymbol{x}^{H}]$ in the multi-beam NOMA network can be expressed as
\begin{equation}
	\label{eq:appA-multibeam-covariance}
	\begin{split}
		\mathbb{E}[\boldsymbol{x}\boldsymbol{x}^{H}]
		&\!=\!\mathbb{E}\left[\left(\sum_{k=1}^{K}\boldsymbol{w}_ks_k+\boldsymbol{\varrho}_t\right)
		\left(\sum_{k=1}^{K}\boldsymbol{w}_ks_k+\boldsymbol{\varrho}_t\right)^{H}\right]\\
		&\!=\!\mathbb{E}\left[\bold{W}^{H}\boldsymbol{s}\boldsymbol{s}^{H}\bold{W}
		\!+\!\boldsymbol{\varrho}_t\boldsymbol{s}^{H}\bold{W}
		\!+\!\bold{W}^{H}\boldsymbol{s}\boldsymbol{\varrho}_t^{H}
		\!+\!\boldsymbol{\varrho}_t\boldsymbol{\varrho}_t^{H}\right],
	\end{split}
\end{equation}
where $\bold{W}=[\boldsymbol{w}_1,\boldsymbol{w}_2,\ldots,\boldsymbol{w}_K]$.

According to Section II-C, the following result can be obtained.
\begin{equation}
	\label{eq:appA-covariance-result}
	\begin{split}
		&\mathbb{E}\left[\bold{W}^{H}\boldsymbol{s}\boldsymbol{s}^{H}\bold{W}
		+\boldsymbol{\varrho}_t\boldsymbol{s}^{H}\bold{W}
		+\bold{W}^{H}\boldsymbol{s}\boldsymbol{\varrho}_t^{H}
		+\boldsymbol{\varrho}_t\boldsymbol{\varrho}_t^{H}\right]\\
		&\qquad=\bold{W}^{H}\bold{W}+\kappa_t\widetilde{{\rm diag}}(\bold{W}^{H}\bold{W}).
	\end{split}
\end{equation}

Therefore, the close-form expression of $\mathbb{E}[|\hat{y}_k|^2]$ in the multi-beam NOMA network can be obtained as follows
\begin{equation}
	\label{eq:appA-multibeam-distortion}
	\begin{split}
		\mathbb{E}[|\hat{y}_k|^2]
		\!=&(\boldsymbol{v}^{H}\bold{H}_k \!+\!\boldsymbol{h}_k^{H})
		\!\!\left(\sum_{k=1}^{K}\boldsymbol{w}_k\boldsymbol{w}_k^{H}
		\!+\!\kappa_t\widetilde{{\rm diag}}\left(\sum_{k=1}^{K}\boldsymbol{w}_k\boldsymbol{w}_k^{H}\right)\!\!\right)\\
		&\times(\boldsymbol{h}_k \!+\!\bold{H}_k^{H}\boldsymbol{v})\!+\!\sigma_k^2,
		\enspace \forall k\in\mathcal{K}.
	\end{split}
\end{equation}

Similarly, the close-form expression of $\mathbb{E}[|\hat{y}_k|^2]$ in the single-beam NOMA network can be obtained as follows
\begin{equation}
	\label{eq:appA-singlebeam-distortion}
	\begin{split}
		\mathbb{E}[|\hat{y}_k|^2]
		=&(\boldsymbol{v}^{H}\bold{H}_k+\boldsymbol{h}_k^{H})
		\left(\boldsymbol{w}_c\boldsymbol{w}_c^{H}
		+\kappa_t\widetilde{{\rm diag}}(\boldsymbol{w}_c\boldsymbol{w}_c^{H})\right)\\
		&\times(\boldsymbol{h}_k+\bold{H}_k^{H}\boldsymbol{v})+\sigma_k^2,
		\enspace \forall k\in\mathcal{K}.
	\end{split}
\end{equation}

The proof of \textbf{Proposition 1} is concluded. \hfill$\blacksquare$

\section{Proof of Theorem 1}

According to $\gamma_l^k\geq\gamma_k^k$, we can obtain

\begin{equation}
	\label{eq:appB-sic-inequality}
	\begin{aligned}
		&\left|(\boldsymbol{v}^{H}\bold{H}_k
		\!+\! \boldsymbol{h}_k^{H})\boldsymbol{w}_c\right|^2\rho_k
		\Bigg\{
		(\boldsymbol{v}^{H}\bold{H}_l \!+\! \boldsymbol{h}_l^{H})
		\left(
		\sum_{i=k \!+\! 1}^{K}
		\rho_i\boldsymbol{w}_c\boldsymbol{w}_c^{H}
		\!+\! \bold{\Psi}_c
		\right)
		\\
		&\times
		(\boldsymbol{h}_l \!+\! \bold{H}_l^{H}\boldsymbol{v})
		\!+\! (1 \!+\! \kappa_r)\sigma_l^2
		\Bigg\}
		\leq
		\left|(\boldsymbol{v}^{H}\bold{H}_l
		\!+\! \boldsymbol{h}_l^{H})\boldsymbol{w}_c\right|^2\rho_k\\
		&\Bigg\{\!\!
		(\boldsymbol{v}^{H}\bold{H}_k \!+\! \boldsymbol{h}_k^{H})\!\!
		\left(
		\sum_{i=k \!+\! 1}^{K}
		\rho_i\boldsymbol{w}_c\boldsymbol{w}_c^{H}
		\!+\! \bold{\Psi}_c \!\!
		\right)\!\!
		(\boldsymbol{h}_k \!+\! \bold{H}_k^{H}\boldsymbol{v})
		\!+\! (1 \!+\! \kappa_r)\sigma_k^2 \!
		\Bigg\}.
	\end{aligned}
\end{equation}

The product of the first terms on both sides of the inequality is equal, so we can obtain
\begin{equation}
	\label{eq:appB-channel-order}
	\begin{split}
		&\frac{\left|(\boldsymbol{v}^{H}\bold{H}_k \!+\! \boldsymbol{h}_k^{H})\boldsymbol{w}_c\right|^2\rho_k}
		{(1 \!+\!\kappa_r)\kappa_t(\boldsymbol{v}^{H}\bold{H}_k \!+\!\boldsymbol{h}_k^{H})
			\widetilde{{\rm diag}}(\boldsymbol{w}_c\boldsymbol{w}_c^{H})
			(\boldsymbol{h}_k \!+\!\bold{H}_k^{H}\boldsymbol{v}) \!+\!(1 \!+\!\kappa_r)\sigma_k^2}\\
		&\!\!\leq\!\!
		\frac{\left|(\boldsymbol{v}^{H}\bold{H}_l \!+\!\boldsymbol{h}_l^{H})\boldsymbol{w}_c\right|^2\rho_k}
		{(1 \!+\!\kappa_r)\kappa_t(\boldsymbol{v}^{H}\bold{H}_l \!+\!\boldsymbol{h}_l^{H})
			\widetilde{{\rm diag}}(\boldsymbol{w}_c\boldsymbol{w}_c^{H})
			(\boldsymbol{h}_l \!+\!\bold{H}_l^{H}\boldsymbol{v}) \!+\!(1 \!+\!\kappa_r)\sigma_l^2},
	\end{split}
\end{equation}

\noindent where $\forall k\in\mathcal{S},\forall l\in\mathcal{T}.$

Then, the equivalent-combined channel gain is defined as
$G_k=\frac{|(\boldsymbol{v}^{H}\bold{H}_k+\boldsymbol{h}_k^{H})\boldsymbol{w}_c|^2\rho_k}{\bold{\Psi}_c^e}$,
where $\bold{\Psi}_c^e=(1+\kappa_r)\kappa_t(\boldsymbol{v}^{H}\bold{H}_k+\boldsymbol{h}_k^{H})
\widetilde{{\rm diag}}(\boldsymbol{w}_c\boldsymbol{w}_c^{H})(\boldsymbol{h}_k+\bold{H}_k^{H}\boldsymbol{v})
+(1+\kappa_r)\sigma_k^2$.
The proof of that $G_1\leq G_2\leq\cdots\leq G_K$ can guarantee the satisfaction of the perfect SIC condition $\gamma_l^k\geq\gamma_k^k$ is the inverse of the above process.

The proof of \textbf{Theorem 1} is concluded. \hfill$\blacksquare$

\section{Proof of Proposition 2}

According to Section II-B, the first part of inequality~\eqref{Pr_2} can be rewritten as
\begin{equation}
	\label{eq:appC-expansion}
	\begin{split}
		&(\boldsymbol{v}^{H}\bold{H}_l+\boldsymbol{h}_l^{H})\bold{\Phi}_k
		(\boldsymbol{h}_l+\bold{H}_l^{H}\boldsymbol{v})\\
		=&(\boldsymbol{v}^{H}(\overline{\bold{H}}_l+\Delta\bold{H}_l)+\boldsymbol{h}_l^{H})\bold{\Phi}_k
		(\boldsymbol{h}_l+(\overline{\bold{H}}_l+\Delta\bold{H}_l)^{H}\boldsymbol{v})\\
		=&{\rm Tr}\left\{\!(\boldsymbol{v}^{H}\overline{\bold{H}}_l \!+\!\boldsymbol{h}_l^{H})\bold{\Phi}_k
		(\boldsymbol{h}_l \!+\!\overline{\bold{H}}_l^{H}\boldsymbol{v})
		\!+\!(\boldsymbol{v}^{H}\overline{\bold{H}}_l \!+\!\boldsymbol{h}_l^{H})\bold{\Phi}_k\Delta\bold{H}_l^{H}\boldsymbol{v}\right.\\
		&\left.+\boldsymbol{v}^{H}\Delta\bold{H}_l\bold{\Phi}_k
		(\boldsymbol{h}_l+\overline{\bold{H}}_l^{H}\boldsymbol{v})
		+\boldsymbol{v}^{H}\Delta\bold{H}_l\bold{\Phi}_k\Delta\bold{H}_l^{H}\boldsymbol{v}\right\},
	\end{split}
\end{equation}
\noindent where $\forall k\in\mathcal{K},\forall l\in\mathcal{T}.$

Defining $\bold{V}=\boldsymbol{v}\boldsymbol{v}^{H}$, based on matrix properties \cite{ref37,ref38}, the following expression can be obtained.
\begin{equation}
	\label{eq:appC-quadratic-term}
	\begin{split}
		{\rm Tr}\left\{\boldsymbol{v}^{H}\Delta\bold{H}_l\bold{\Phi}_k\Delta\bold{H}_l^{H}\boldsymbol{v}\right\}
		&\!=\!{\rm vec}^{H}(\Delta\bold{H}_l)(\bold{\Phi}_k^{H}\!\otimes\!\bold{V}){\rm vec}(\Delta\bold{H}_l)\\
		&\!=\!\phi_{H,l}^{2}\bold{\Lambda}_{H,l}^{H}(\bold{\Phi}_k^{H}\otimes\bold{V})\bold{\Lambda}_{H,l}.
	\end{split}
\end{equation}

Similarly, the rest of the terms in~\eqref{eq:appC-expansion} can be rewritten as
\begin{equation}
	\label{eq:appC-linear-term}
	\begin{split}
		&{\rm Tr}\left\{(\boldsymbol{v}^{H}\overline{\bold{H}}_l+\boldsymbol{h}_l^{H})\bold{\Phi}_k\Delta\bold{H}_l^{H}\boldsymbol{v}\right\}\\
		=&{\rm Tr}\left\{\boldsymbol{v}^{H}\Delta\bold{H}_l\bold{\Phi}_k
		(\boldsymbol{h}_l+\overline{\bold{H}}_l^{H}\boldsymbol{v})\right\}\\
		=&{\rm vec}^{H}(\Delta\bold{H}_l){\rm vec}\left((\bold{V}\overline{\bold{H}}_l+\boldsymbol{v}\boldsymbol{h}_l^{H})\bold{\Phi}_k\right)\\
		=&\phi_{H,l}\bold{\Lambda}_{H,l}^{H}{\rm vec}\left((\bold{V}\overline{\bold{H}}_l+\boldsymbol{v}\boldsymbol{h}_l^{H})\bold{\Phi}_k\right),
	    \forall k\in\mathcal{K},\forall l\in\mathcal{T}.
	\end{split}
\end{equation}

Substituting~\eqref{eq:appC-expansion}, \eqref{eq:appC-quadratic-term}, and~\eqref{eq:appC-linear-term} into~\eqref{Pr_2} yields
\begin{equation}
	\label{eq:appC-result}
	\begin{split}
		{\rm Pr}&\Big\{\phi_{H,l}^{2}\bold{\Lambda}_{H,l}^{H}(\bold{\Phi}_k^{H}\!\!\otimes\!\!\bold{V})\bold{\Lambda}_{H,l}
		\!+\!(\boldsymbol{v}^{H}\overline{\bold{H}}_l+\boldsymbol{h}_l^{H})\bold{\Phi}_k
		(\boldsymbol{h}_l+\overline{\bold{H}}_l^{H}\boldsymbol{v})\\
		&\!+\!2{\rm Re}\left\{\!\phi_{H,l}\bold{\Lambda}_{H,l}^{H}
		{\rm vec}\!\left(\!(\bold{V}\overline{\bold{H}}_l \!+\!\boldsymbol{v}\boldsymbol{h}_l^{H}\!)\bold{\Phi}_k \!\right)\!\right\}
		\!-\!(1 \!+\!\kappa_r)\sigma_k^2\!\geq0\Big\}\\
		&\!\geq\!1 \!-\! P_{out}^{k},\enspace \forall k\in\mathcal{K},\forall l\in\mathcal{T}.
	\end{split}
\end{equation}

The proof of \textbf{Proposition 2} is concluded. \hfill$\blacksquare$

\section{Proof of Proposition 3}

According to Section II-B, the first part of inequality~\eqref{Pr_2} can be rewritten as
\begin{equation}
	\label{eq:appD-expansion}
	\begin{split}
		&(\boldsymbol{v}^{H}\bold{H}_l+\boldsymbol{h}_l^{H})\bold{\Phi}_k
		(\boldsymbol{h}_l+\bold{H}_l^{H}\boldsymbol{v})\\
		=&\left[(\boldsymbol{v}^{H}\overline{\bold{H}}_l+\overline{\boldsymbol{h}}_l^{H})
		+(\boldsymbol{v}^{H}\Delta\bold{H}_l+\Delta\boldsymbol{h}_l^{H})\right]\bold{\Phi}_k\\
		&\times	\left[(\overline{\boldsymbol{h}}_l+\overline{\bold{H}}_l^{H}\boldsymbol{v})
		+(\Delta\boldsymbol{h}_l+\Delta\bold{H}_l^{H}\boldsymbol{v})\right]\\
		=&(\boldsymbol{v}^{H}\overline{\bold{H}}_l+\overline{\boldsymbol{h}}_l^{H})\bold{\Phi}_k
		(\overline{\boldsymbol{h}}_l+\overline{\bold{H}}_l^{H}\boldsymbol{v})\\
		&+(\boldsymbol{v}^{H}\overline{\bold{H}}_l+\overline{\boldsymbol{h}}_l^{H})\bold{\Phi}_k
		(\Delta\boldsymbol{h}_l+\Delta\bold{H}_l^{H}\boldsymbol{v})\\
		&+(\boldsymbol{v}^{H}\Delta\bold{H}_l+\Delta\boldsymbol{h}_l^{H})\bold{\Phi}_k
		(\overline{\boldsymbol{h}}_l+\overline{\bold{H}}_l^{H}\boldsymbol{v})\\
		&+(\boldsymbol{v}^{H}\Delta\bold{H}_l+\Delta\boldsymbol{h}_l^{H})\bold{\Phi}_k
		(\Delta\boldsymbol{h}_l+\Delta\bold{H}_l^{H}\boldsymbol{v}),
	\end{split}
\end{equation}
\noindent where $\enspace \forall k\in\mathcal{K},\forall l\in\mathcal{T}.$

Defining
$\bold{V}=\boldsymbol{v}\boldsymbol{v}^{H}$ and $\overline{\boldsymbol{g}}_l^{H}
\triangleq	\boldsymbol{v}^{H}\overline{\bold{H}}_l
+\overline{\boldsymbol{h}}_l^{H},$
based on the matrix properties in~\cite{ref37,ref38},
the following expressions can be obtained.
\begin{equation}
	\label{eq:appD-linear-term}
	\begin{split}
		&2{\rm Re}\left\{
		\overline{\boldsymbol{g}}_l^{H}
		\bold{\Phi}_k
		(\Delta\boldsymbol{h}_l
		+\Delta\bold{H}_l^{H}\boldsymbol{v})
		\right\}
		\\
		={}&2{\rm Re}\left\{
		\overline{\boldsymbol{g}}_l^{H}
		\bold{\Phi}_k\Delta\boldsymbol{h}_l
		+
		{\rm vec}^{T}\left(
		\boldsymbol{v}
		\overline{\boldsymbol{g}}_l^{H}
		\bold{\Phi}_k
		\right)
		{\rm vec}(\Delta\bold{H}_l^{*})
		\right\}
		\\
		={}&2{\rm Re}\left\{
		\phi_{h,l}
		\overline{\boldsymbol{g}}_l^{H}
		\bold{\Phi}_k
		\bold{\Lambda}_{h,l}
		+
		\phi_{H,l}
		{\rm vec}^{T}\left(
		\boldsymbol{v}
		\overline{\boldsymbol{g}}_l^{H}
		\bold{\Phi}_k
		\right)
		\bold{\Lambda}_{H,l}^{*}
		\right\}
		\\
		={}&2{\rm Re}\left\{
		\widetilde{\boldsymbol{r}}_{l,k}^{H}
		\widetilde{\bold{\Lambda}}_l
		\right\},
		\quad
		\forall k\in\mathcal{K},
		\forall l\in\mathcal{T}.
	\end{split}
\end{equation}

and
\begin{equation}
	\label{eq:appD-quadratic-term}
	\begin{split}
		&(\boldsymbol{v}^{H}\Delta\bold{H}_l+\Delta\boldsymbol{h}_l^{H})\bold{\Phi}_k
		(\Delta\boldsymbol{h}_l+\Delta\bold{H}_l^{H}\boldsymbol{v})\\
		=\ &\phi_{h,l}^{2}\bold{\Lambda}_{h,l}^{H}\bold{\Phi}_k\bold{\Lambda}_{h,l}
		+2{\rm Re}\left\{\phi_{h,l}\phi_{H,l}\bold{\Lambda}_{h,l}^{H}
		(\bold{\Phi}_k\otimes\boldsymbol{v}^{T})\bold{\Lambda}_{H,l}^{*}\right\}\\
		&+\phi_{H,l}^{2}\bold{\Lambda}_{H,l}^{T}(\bold{\Phi}_k\otimes\bold{V}^{T})\bold{\Lambda}_{H,l}^{*}\\
		=\ &\widetilde{\bold{\Lambda}}_l^{H}\widetilde{\bold{Q}}_{l,k}\widetilde{\bold{\Lambda}}_l.
	\end{split}
\end{equation}

Substituting~\eqref{eq:appD-expansion}, \eqref{eq:appD-linear-term}, and~\eqref{eq:appD-quadratic-term} into~\eqref{Pr_2} yields
\begin{equation}
	\label{eq:appD-result}
	\begin{split}
		{\rm Pr}\left\{\widetilde{\bold{\Lambda}}_l^{H}\widetilde{\bold{Q}}_{l,k}\widetilde{\bold{\Lambda}}_l
		\!+\!2{\rm Re}\left\{\widetilde{\boldsymbol{r}}_{l,k}^{H}\widetilde{\bold{\Lambda}}_l\right\}
		\!+\!\widetilde{c}_{l,k}\!\geq\!0\right\}
		\!\geq\!1\!-\! P_{out}^{k}.
	\end{split}
\end{equation}

The proof of \textbf{Proposition 3} is concluded. \hfill$\blacksquare$

\bibliographystyle{IEEEtran}
\bibliography{myref}

\end{document}